\documentclass[preprint,aps,showpacs,nofootinbib,preprintnumbers,amsmath,amssymb,superscriptaddress]{revtex4-1}
\usepackage{amssymb}
\usepackage{epsfig}
\usepackage{latexsym}
\usepackage{graphicx}
\usepackage{subfigure}
\usepackage{dcolumn}
\usepackage{bm}
\usepackage[usenames ,dvipsnames]{xcolor}
\usepackage[utf8]{inputenc}
\usepackage{slashed}
\usepackage{cleveref}

\begin{document}
\title{Muon $g-2$ Anomaly from a Massive Spin-2 Particle}
\author{Da~Huang\footnote{dahuang@bao.ac.cn}}
\affiliation{National Astronomical Observatories, Chinese Academy of Sciences, Beijing, 100012, China}
\affiliation{School of Fundamental Physics and Mathematical Sciences, Hangzhou Institute for Advanced Study, UCAS, Hangzhou 310024, China}
\affiliation{International Centre for Theoretical Physics Asia-Pacific, Beijing/Hangzhou, China}
\author{Chao-Qiang~Geng\footnote{geng@phys.nthu.edu.tw}}
\affiliation{School of Fundamental Physics and Mathematical Sciences, Hangzhou Institute for Advanced Study, UCAS, Hangzhou 310024, China}
\affiliation{International Centre for Theoretical Physics Asia-Pacific, Beijing/Hangzhou, China}
\author{Jiajun~Wu\footnote{wujiajun@itp.ac.cn}}
\affiliation{School of Fundamental Physics and Mathematical Sciences, Hangzhou Institute for Advanced Study, UCAS, Hangzhou 310024, China}
\affiliation{International Centre for Theoretical Physics Asia-Pacific, Beijing/Hangzhou, China}

\date{\today}
\begin{abstract}
We investigate the possibility to interpret the muon $g-2$ anomaly in terms of a massive spin-2 particle, $G$, which can be identified as the first Kaluza-Klein graviton in the generalized Randall-Sundrum model. In particular, we obtain the leading-order contributions to the muon $g-2$ by calculating the relevant one-loop Feynman diagrams induced by $G$. The analytic expression is shown to keep the gauge invariance of the quantum electrodynamics and to be consistent with the expected UV divergence structure. Moreover, we impose the theoretical bounds from the perturbativity {and the experimental constraints from LHC and LEP-II on} our model. Especially, we derive  novel perturbativity constraints on nonrenormalizable operators related to $G$, which are the natural generalization of the counterpart for the renormalizable operators. As a result, we show that there exists a substantial parameter space, which can accommodate the muon $g-2$ anomaly {allowed by all constraints}. Finally, we also make comments on the possible explanation of the electron $g-2$ anomalies with the massive spin-2 particle.      
\end{abstract}

\maketitle

\section{Introduction}\label{s1}
The long-standing discrepancy between the measurement of the muon magnetic dipole moment $(g-2)_\mu$ and the Standard Model (SM) prediction is one of the greatest puzzles in  particle physics~\cite{pdg}, which might be the hint to  new physics beyond the SM. This problem becomes more severe recently since the Muon $g-2$ Collaboration at Fermilab has reported a new measurement of the muon magnetic moment $a_\mu \equiv (g-2)_\mu/2$ with the result given by~\cite{Muong-2:2021ojo}
\begin{eqnarray}
     a_\mu^{\rm FNAL} = (116592040\pm 54)\times 10^{-11}\,.
\end{eqnarray} 
When combining the earlier data from the experiment at Brookhaven~\cite{Muong-2:2006rrc}, the anomalous contribution to $(g-2)_\mu$ is given by
\begin{eqnarray}\label{ExpG2}
	\Delta a_\mu = a^{\rm Exp}_\mu - a^{\rm SM}_\mu = (251\pm 59)\times 10^{-11}\,,
\end{eqnarray}
in which {the latest SM prediction obtained by combining various contributions~\cite{Aoyama:2012wk,Aoyama:2019ryr,Czarnecki:2002nt,Gnendiger:2013pva,Davier:2017zfy,Keshavarzi:2018mgv,Colangelo:2018mtw,Hoferichter:2019mqg,Davier:2019can,Keshavarzi:2019abf,Kurz:2014wya,Melnikov:2003xd,Masjuan:2017tvw,Colangelo:2017fiz,Hoferichter:2018kwz,Gerardin:2019vio,Bijnens:2019ghy,Colangelo:2019uex,Blum:2019ugy,Colangelo:2014qya} is $a_\mu^{\rm SM} = (116591810\pm 43)\times 10^{-11}$ (see {\it e.g.}, Ref.~\cite{Aoyama:2020ynm} for a recent review). More recently, there are several lattice QCD results~\cite{Borsanyi:2020mff, Alexandrou:2022amy, Ce:2022kxy, FermilabLattice:2022smb} on the hadronic vacuum polarization contribution to the muon $g-2$, which indicate that the discrepancy might be weakened to below $4\sigma$. Nevertheless, it is still of great importance to take the muon $g-2$ anomaly seriously.  }

Apart from the strong evidence to muon $g-2$ anomaly, the latest measurement of the fine-structure constant $\alpha$ would also imply a discrepancy between the SM calculation and the experimental measurement of the electron $g-2$. In the literature, there have been two recent measurements of $\alpha$ from Laboratoire Kastler Brossel (LKB) with $^{87}{\rm Rb}$ atoms~\cite{Morel:2020dww} and at Berkeley with $^{137}{\rm Cs}$ atoms~\cite{Parker:2018vye}, which lead to the following SM predictions~\cite{Aoyama:2012wj,Aoyama:2019ryr} for the anomalous electron $(g-2)_e$
\begin{eqnarray}
	\Delta a_e^{\rm LKB} &=& a_e^{\rm exp} - a_e^{\rm LKB} = (4.8\pm 3.0) \times 10^{-13}\,,\nonumber\\
	\Delta a_e^{\rm B} &=& a_e^{\rm exp} - a_e^{\rm B} = (-8.8 \pm 3.6) \times 10^{-13}\,,
\end{eqnarray}
and their deviations of theoretical values from the experimental result $a_e^{\rm exp}$~\cite{Hanneke:2008tm} are at $1.6\sigma$ and $-2.4 \sigma$, respectively. It is interesting to have a common explanation to both electron and muon $g-2$ data in one single framework.   

In the literature, there have already been many attempts to interpret the muon $g-2$ anomaly in terms of various models beyond the SM (for a recent review see {\it e.g.}~\cite{Athron:2021iuf} and references therein). In the present paper, we explore an alternative explanation to the $(g-2)_{\mu,\, e}$ anomalies which are induced by a new massive spin-2 particle $G$~\cite{Graesser:1999yg,Kang:2020huh,Hong:2017tel}. Note that $G$ can naturally arise as the Kaluza-Klein (KK) graviton in the five-dimensional Randall-Sundrum (RS) model~\cite{Randall:1999ee}, which is motivated to solve the gauge hierarchy problem in the SM. It is remarkable to note that the massive spin-2 particle can couple to the SM particles nonuniversally in the generalized RS models~\cite{Davoudiasl:1999tf,Pomarol:1999ad,Chang:1999nh,Davoudiasl:2000wi,Batell:2005wa,Batell:2006dp,Fok:2012zk,Lee:2013bua,Han:2015cty,Geng:2016xin,Falkowski:2016glr,Dillon:2016fgw,Dillon:2016tqp,Kraml:2017atm,Geng:2018hpq,Goyal:2019vsw} due to the different localization of SM fields in the extra-dimensional bulk. Especially, we here focus on the spin-2 particle coupling to the SM leptons and photons, which may give rise to novel one-loop contributions to the lepton $g-2$. 

The paper is organized as follows. In Sec.~\ref{model}, we present the effective interactions between the massive spin-2 particle $G$ and the SM photon and leptons. Then the one-loop Feynman diagrams and the final analytic expressions for the lepton anomalous magnetic moment $(g-2)_\ell$ induced by $G$ are presented in Sec.~\ref{G2me}. Sec.~\ref{TheoreticalConstraint} is devoted to the investigation of the theoretical bounds from the perturbativity in our model of the massive spin-2 particle. In Sec.~\ref{SecResult}, we explore numerically the parameter space that can explain the lepton $g-2$ anomalies while satisfying the above theoretical constraints. {Sec.~\ref{Collider} is devoted to the studies of existing collider constraints from LHC and LEP-II. Finally, we conclude in Sec.~\ref{Conclusion} in which a short discussion is given for the lepton-flavor-violating (LFV) and $CP$-violating (CPV) effects.} In Appendix~\ref{AppG2} we present details for calculating various one-loop Feynman diagrams contributing to the lepton $g-2$ in our massive graviton model. Especially, we have checked the gauge invariance of the quantum electrodynamics in the Barr-Zee-type diagrams~\cite{Bjorken:1977vt,Barr:1990vd}.

\section{Lagrangian for the Massive Graviton}\label{model}
We are working in the framework of the effective field theory of the spin-2 particle $G$, with the relevant Lagrangian given by~\cite{Han:1998sg}
\begin{eqnarray}\label{LagG}
    {\cal L}_G = -\frac{1}{\Lambda} G_{\mu\nu} \Big[c_\gamma T_\gamma^{\mu\nu} + \sum_{\ell=e,\mu,\tau} c_\ell T_\ell^{\mu\nu}\Big]\,,
\end{eqnarray}
where $T_\gamma^{\mu\nu}$ and $T_\ell^{\mu\nu}$ represent the energy-momentum tensors of photons and charged leptons $\ell$ defined as follows
\begin{eqnarray}\label{DefEMT}
    T_\ell^{\mu\nu} &=& \frac{i}{4} \bar{\ell} (\gamma^\mu \partial^\nu + \gamma^\nu \partial^\mu) \ell -\frac{i}{4} (\partial^\mu \bar{\ell}\gamma^\nu + \partial^\nu \bar{\ell}\gamma^\mu) \ell \nonumber\\
    && -i\eta^{\mu\nu} [\bar{\ell} \gamma^\rho \partial_\rho \ell + im_\ell \bar{\ell}\ell -\frac{1}{2}\partial^\rho (\bar{\ell}\gamma_\rho \ell)]\,, \nonumber\\
    T_\gamma^{\mu\nu} &=& \frac{1}{4} \eta^{\mu\nu} F^{\rho\sigma}F_{\rho\sigma} - F^{\mu\rho}F^{\nu}_\rho \nonumber\\
    && -\frac{1}{\xi} \left[\eta^{\mu\nu}\left(\partial^\rho \partial^\sigma A_\sigma A_\rho + \frac{1}{2}(\partial^\rho A_\rho)^2\right)-\left(\partial^\mu \partial^\rho A_\rho A^\nu + \partial^\nu \partial^\rho A_\rho A^\mu\right)\right]\,,
\end{eqnarray}
with $\xi$ the gauge parameter for the photon field. This Lagrangian can be easily derived from the generalized RS model with the massive spin-2 particle identified as the first KK excitation of the graviton~\cite{Randall:1999ee,Davoudiasl:1999tf,Pomarol:1999ad,Chang:1999nh,Davoudiasl:2000wi,Falkowski:2016glr}. Traditionally, in order to solve the hierarchy problem, this massive graviton should be strongly coupled to the third-generation quarks and SM gauge bosons. However, we do not show them here and only list terms relevant to our discussion of charged lepton $g-2$ anomalies. 

Moreover, there should be the lepton flavor off-diagonal interactions with the massive spin-2 particle such as
\begin{eqnarray}\label{OffDiagInt}
	-\frac{G_{\mu\nu}}{\Lambda} \left[ c_{\ell^\prime \ell} T_{\ell^\prime \ell}^{\mu\nu} + {\rm H.c.} \right]\,,
\end{eqnarray}
with
\begin{eqnarray}\label{OffDiagET}
T_{\ell^\prime \ell}^{\mu\nu} &\equiv& \frac{i}{4} \bar{\ell}^\prime \left(\gamma^\mu \partial^\nu + \gamma^\nu \partial^\mu\right)\ell - \frac{i}{4} \left( \partial^\mu \bar{\ell}^\prime \gamma^\nu + \partial^\nu \bar{\ell}^\prime \gamma^\mu \right)\ell  \nonumber\\
&& -i\eta^{\mu\nu} \left[ \bar{\ell}^\prime \gamma^\rho \partial_\rho \ell + i m_{\ell^\prime \ell} \bar{\ell}^\prime \ell - \frac{1}{2} \partial^\rho (\bar{\ell}^\prime \gamma_\rho \ell) \right]\,,
\end{eqnarray}
where $\ell^\prime$ and $\ell$ denote different charged lepton flavors and $m_{\ell^\prime \ell}$ is a parameter with unit mass dimension which is determined by the extra-dimensional wavefunctions of the massive graviton and various lepton fields. These interactions would lead to the charged lepton LFV observables, such as $\mu \to e\gamma$~\cite{MEG:2016leq}, $\mu^+ \to e^+ e^+ e^-$~\cite{SINDRUM:1987nra}, $\mu^-$-$e^-$ conversions in nuclei~\cite{SINDRUMII:2006dvw} and so on, which has been stringently constrained  experimentally. On the other hand, due to the non-Hermitian nature of stress-energy tensors $T_{\ell^\prime \ell}$, the coupling coefficients $c_{\ell^\prime \ell}$ should be complex in general, so that they can induce the CPV effects like the electric dipole moments (EDMs) of the electron~\cite{ACME:2018yjb} and muon~\cite{Muong-2:2008ebm}, which should also be strongly suppressed as required by experiments. Hence, given the impressive LFV and CPV constraints on these flavor off-diagonal interactions, it is expected that their contributions to the lepton magnetic dipole moments should be sub-dominant to those induced by the flavor diagonal interactions presented in Eqs.~(\ref{LagG}) and (\ref{DefEMT}). We will come back to this issue later.    

\section{Massive Spin-2 Particle Contributions to Charged Lepton $(g-2)_\ell$}\label{G2me}
\begin{figure}[!htb]
	\centering
	\hspace{-5mm}
	\includegraphics[width=0.32\linewidth]{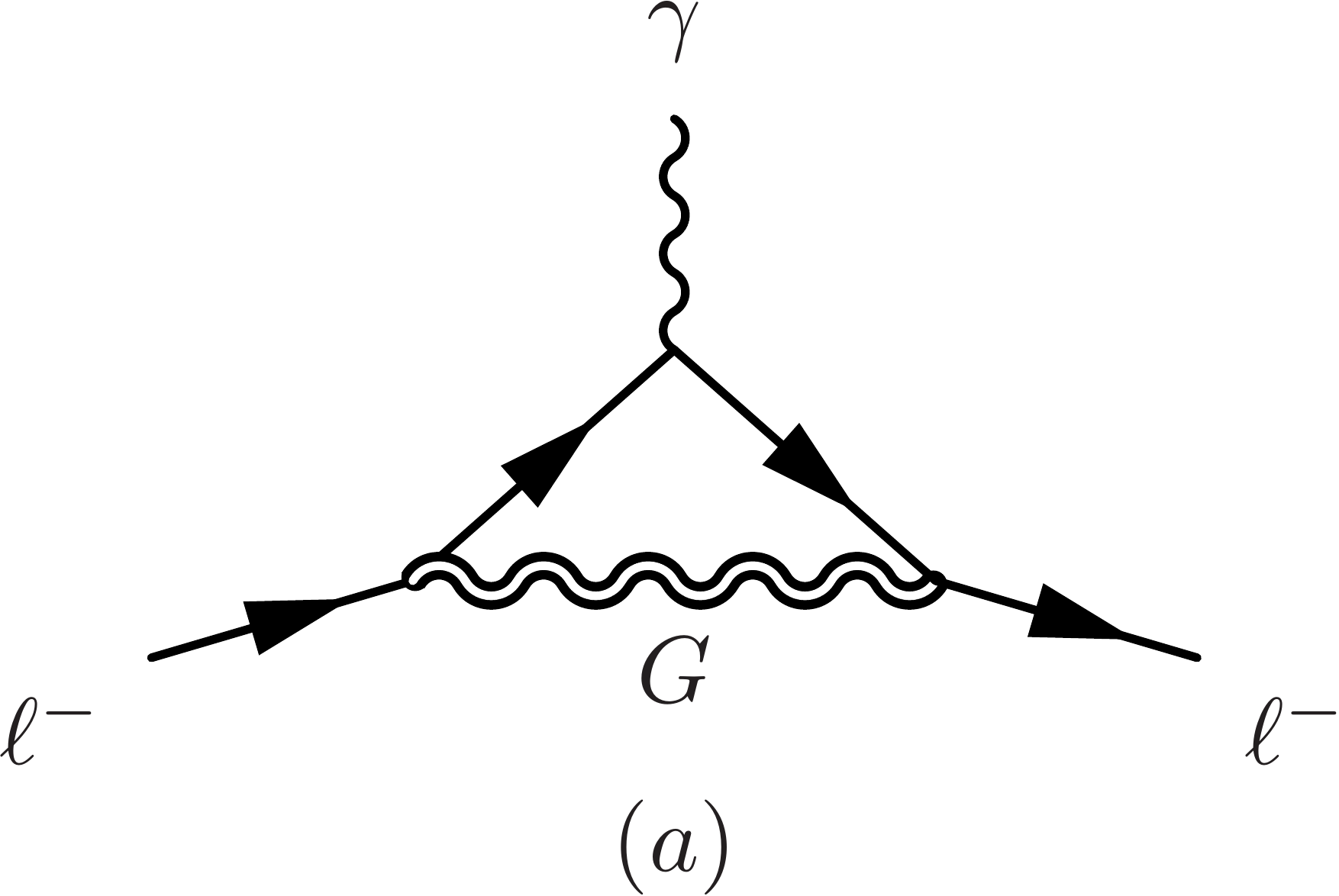}
	\includegraphics[width=0.32\linewidth]{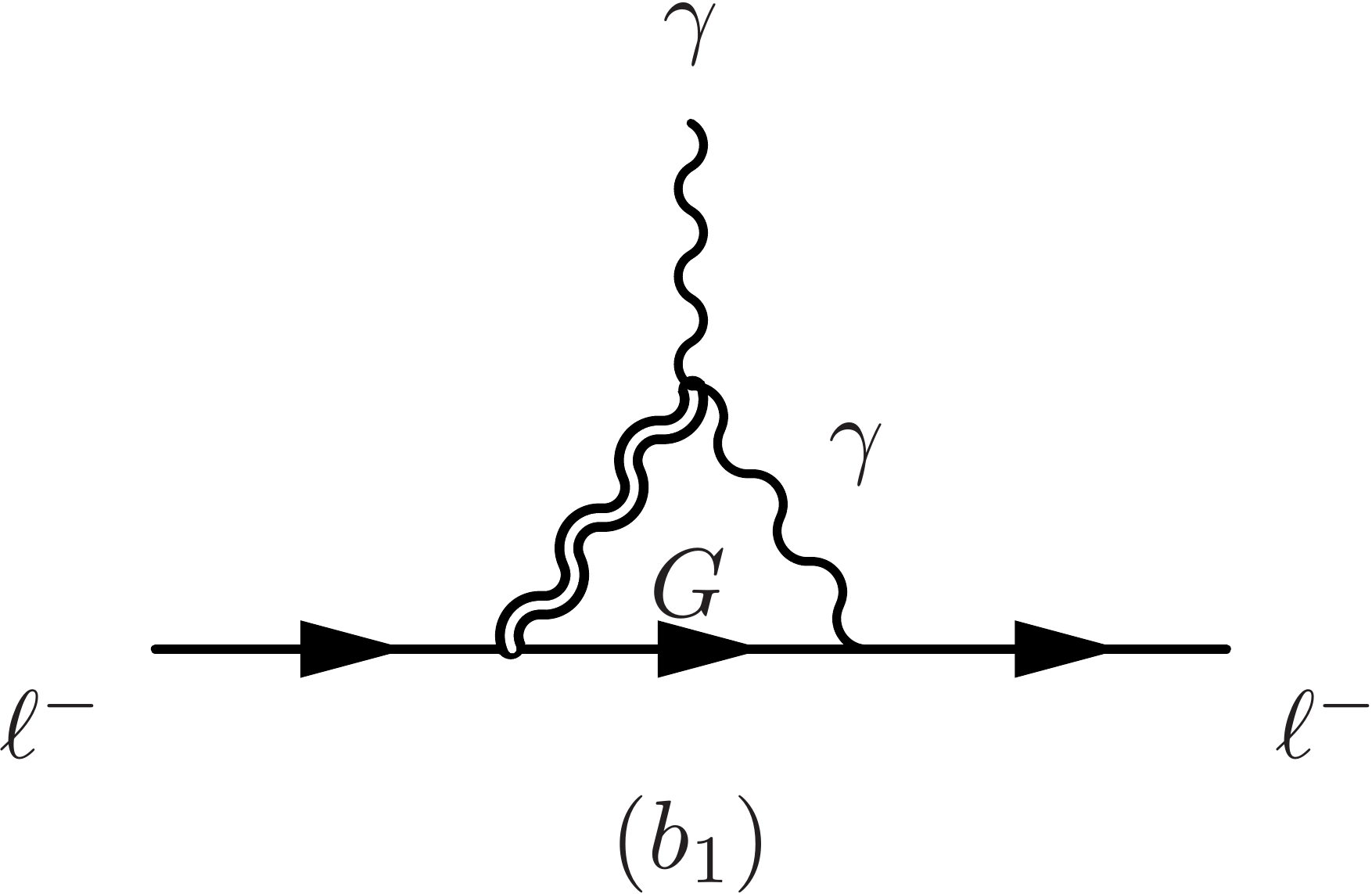}
	\includegraphics[width=0.32\linewidth]{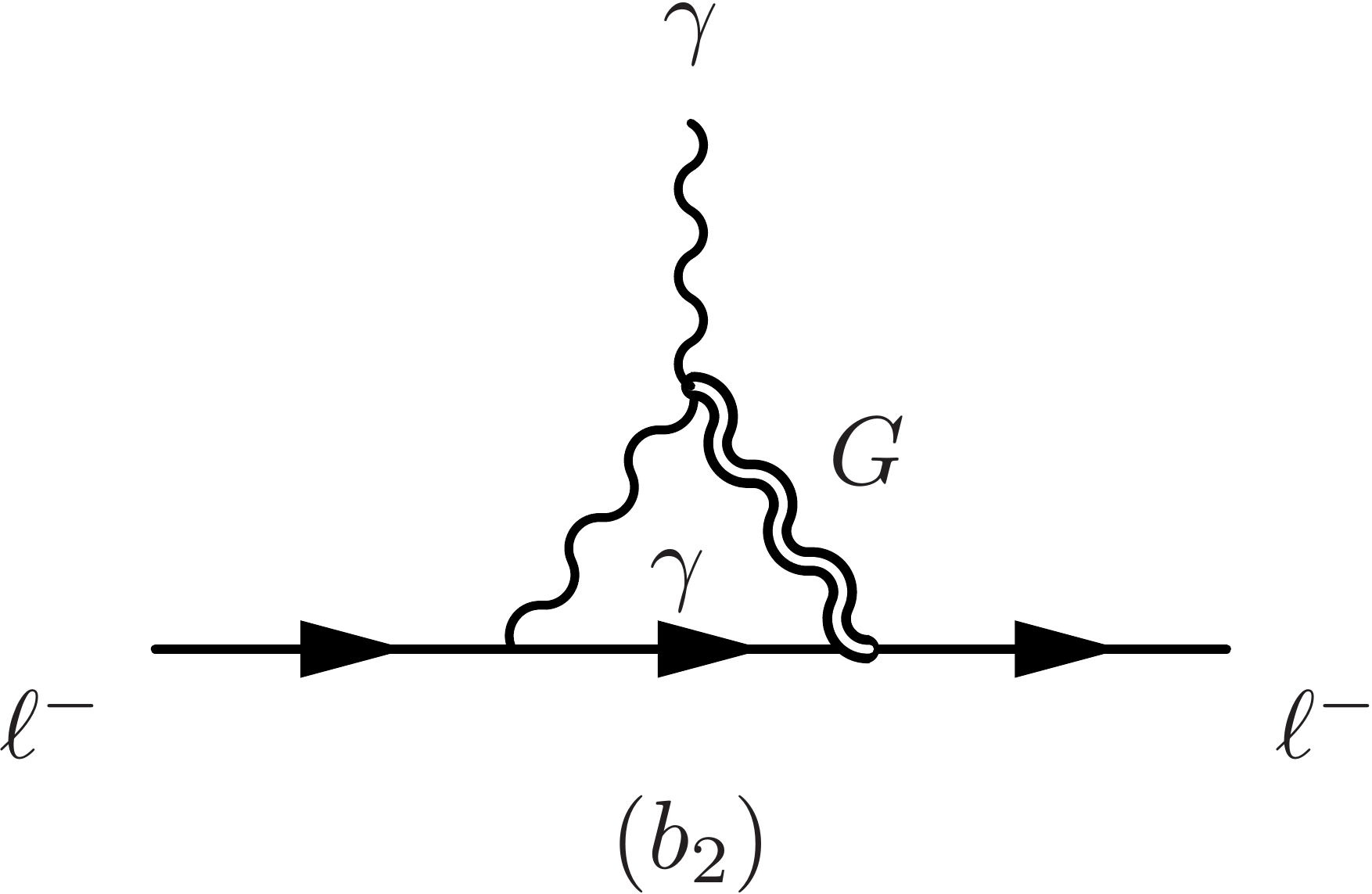}
    \includegraphics[width=0.32\linewidth]{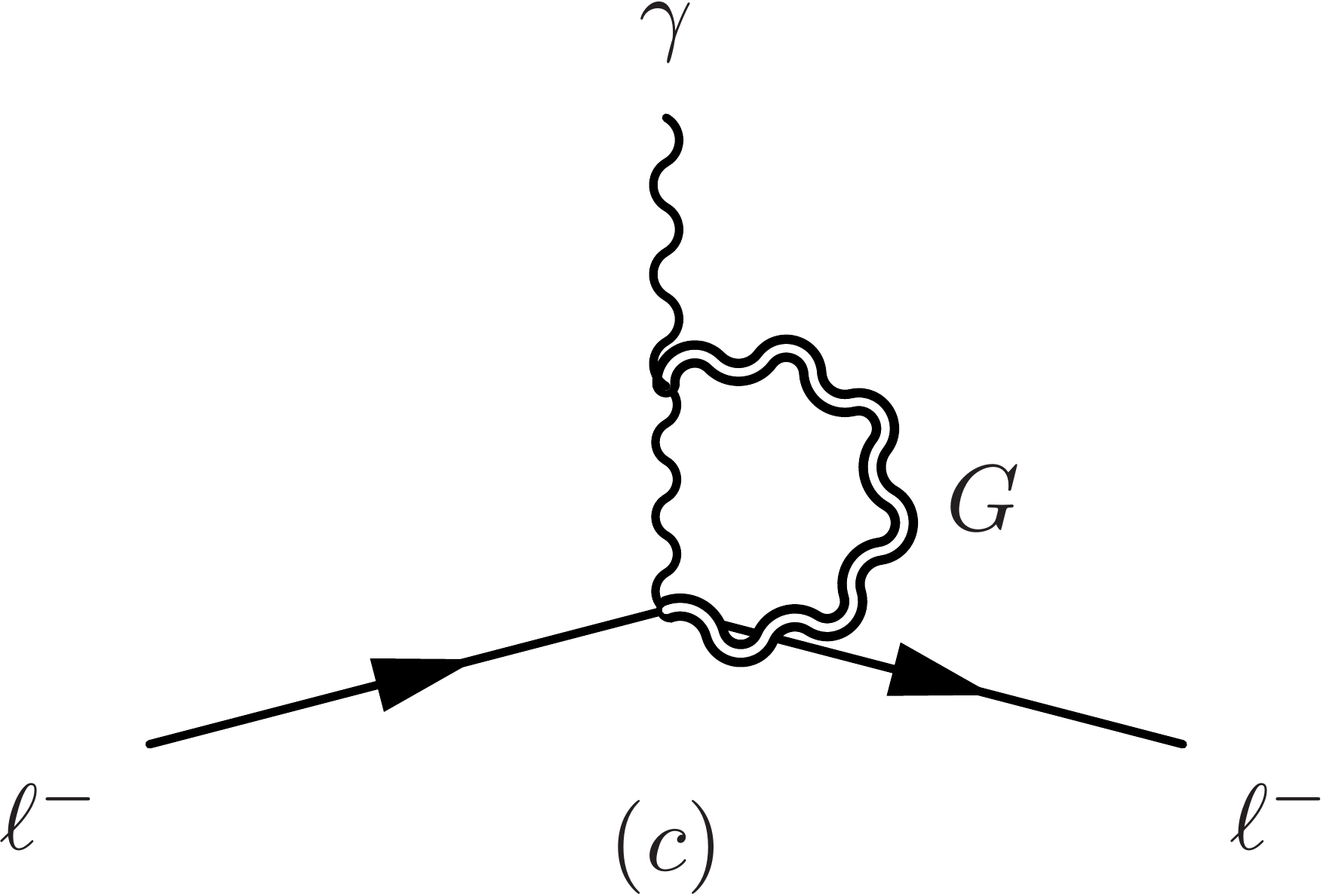}
	\includegraphics[width=0.32\linewidth]{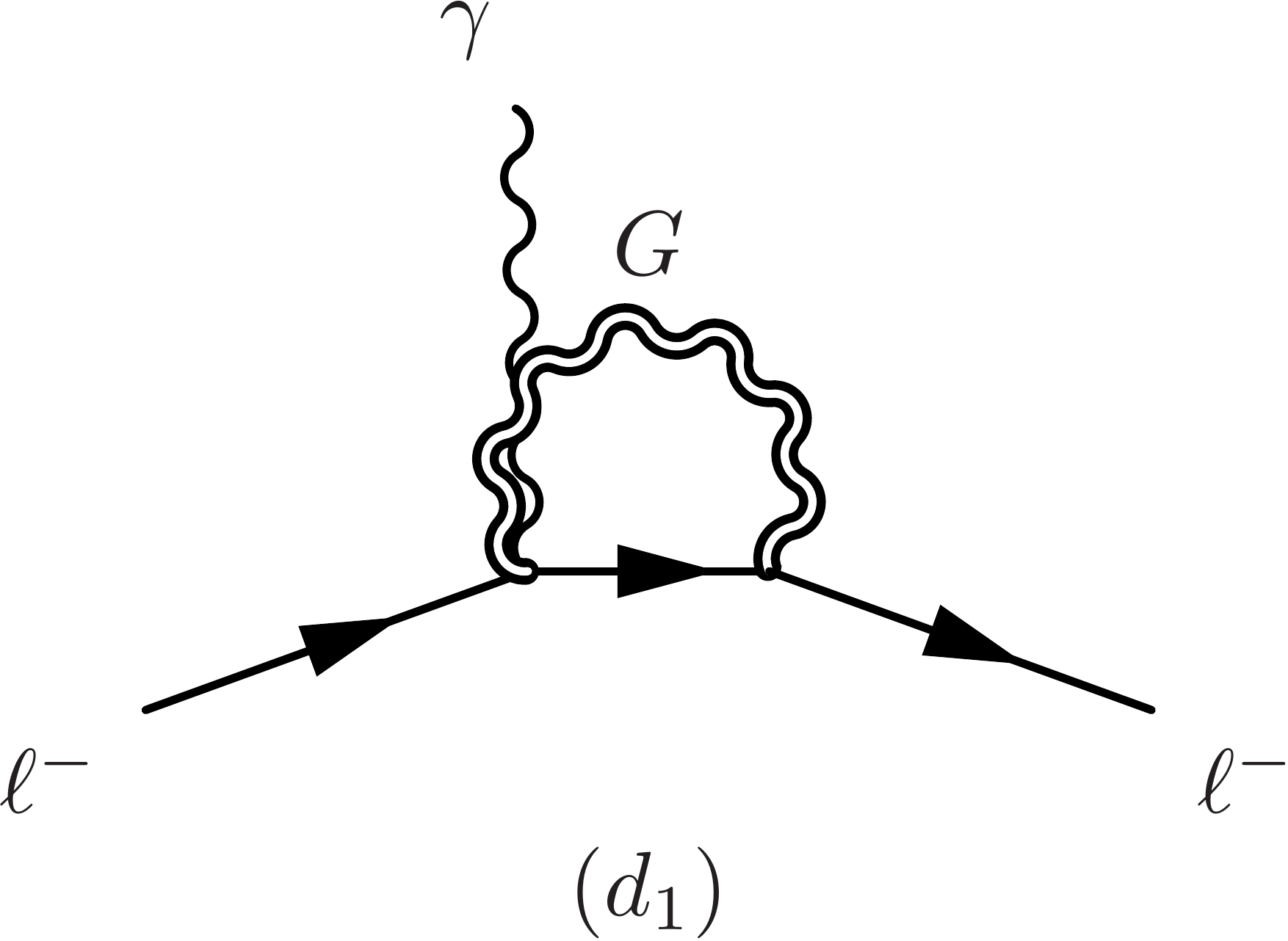}
	\includegraphics[width=0.32\linewidth]{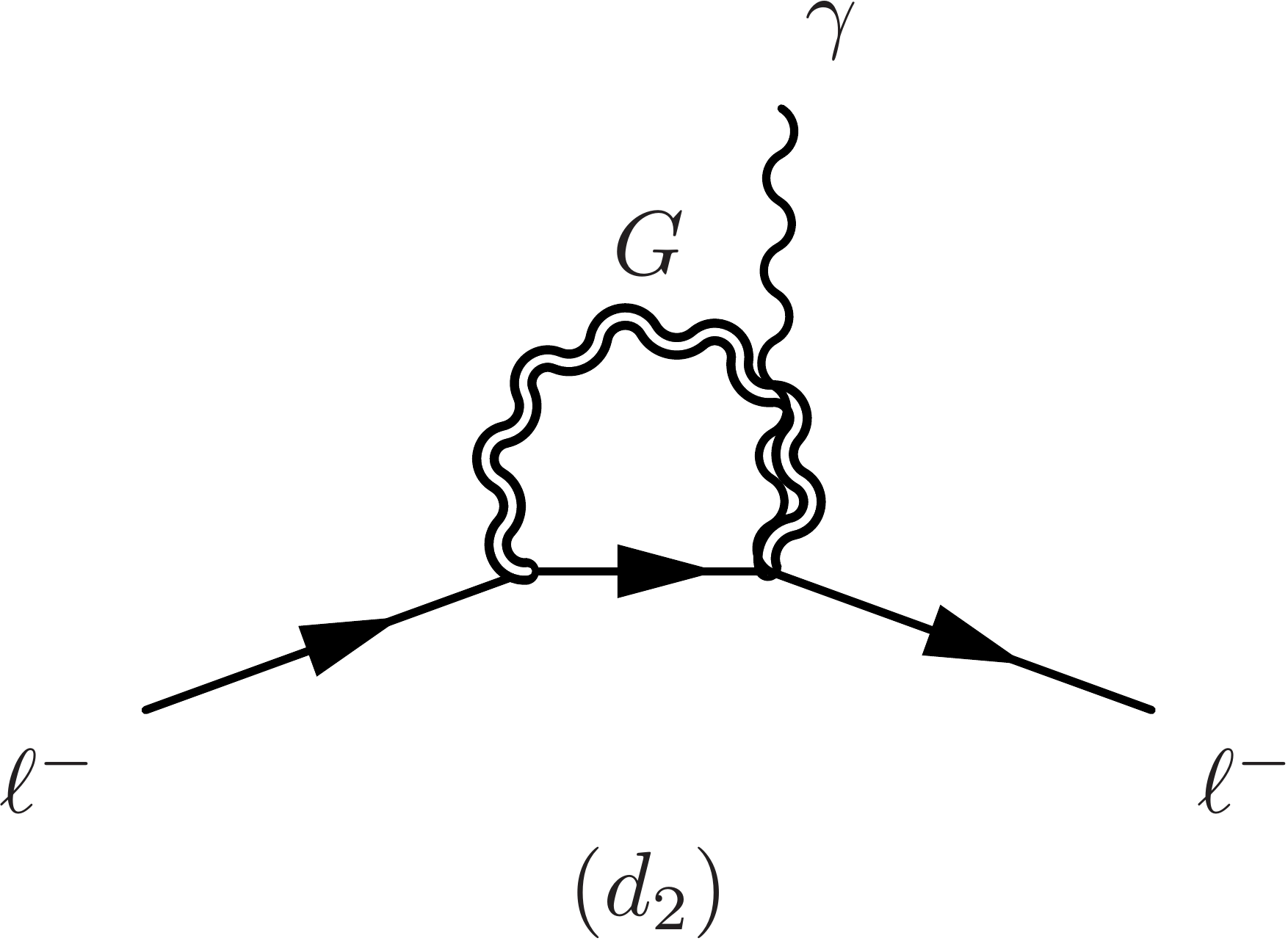}
	\caption{One-loop Feynman diagrams giving rise to the leading-order contributions to the muon $g-2$.}\label{FigG2}
\end{figure}

In this section, we study the massive spin-2 particle interpretation of the possible lepton $(g-2)_{\mu,\,e}$ anomalies. Given the Lagrangian in Eq.~(\ref{LagG}), we can draw one-loop Feynman diagrams shown in Fig.~\ref{FigG2} which give the leading-order contributions to the anomalous lepton magnetic moments.
Since the calculation of these Feynman diagrams is tedious, here we only present the final results but leaving the details in Appendix A. First of all, with the simple power counting rules, it is easy to see that all loop integrals are highly power-law divergent, with the largest divergence of order of sixth power. However, it can be shown that for the contributions to the lepton $g-2$, the leading divergence can be merely of quartic, {\it i.e.} of ${\cal O}(\Lambda^4)$ with $\Lambda$ identified as the UV cutoff scale appearing in the effective action in Eq.~(\ref{LagG}). Furthermore, explicit calculations show that the diagram $(c)$ cannot contribute to lepton magnetic moments, while, up to ${\cal O}(\Lambda^4)$, the result of $\Delta a_{\ell}$ from the diagram $(a)$ vanishes identically. On the other hand, Feynman diagrams $(b_{1,2})$ and $(d_{1,2})$ do give rise to the leading-order nonzero contributions to the charged lepton $(g-2)$ with their total results given by
\begin{eqnarray}\label{G2total}
	\Delta a_{\ell}^G = \left(\frac{m_\ell}{\Lambda}\right)^2 \left(\frac{\Lambda}{m_G}\right)^4 \left(\frac{c_\ell^2}{48\pi^2} - \frac{c_\ell c_\gamma}{24\pi^2}\right)\,.
\end{eqnarray}
Note that, when computing these one-loop Feynman diagrams, one encounters loop integrals with quartic divergence. Also, since the photon is a gauge boson of quantum electrodynamics, it is required that the final result of $(g-2)_\ell$ should be gauge invariant. Here, we have applied the Loop Reguarization~\cite{Wu:2002xa,Wu:2003dd} method to preserve both the correct divergence power and the gauge structure at the same time, which is impossible for the traditional dimensional reguarization~\cite{tHooft:1972tcz}. 

Note that the single massive spin-2 particle contributions to the muon $g-2$ from the same set of Feynman diagrams in Fig.~\ref{FigG2} were calculated in Ref.~\cite{Graesser:1999yg}, where the author found that all of these Feynman diagrams could give rise to the logarithmically divergent expressions. In particular, when the photon and leptons share a universal coupling to the spin-2 field $G$, {\it i.e.}, $c_\ell = c_\gamma$, the total contribution became remarkably finite. It was argued in Ref.~\cite{Graesser:1999yg} that the decrease of the degree of UV divergence was caused by the gravitational Ward identity, so that terms containing two or more $k_\alpha$'s in the numerator of the massive spin-2 field propagator in Eq.~(\ref{PropG}) vanished in these loop calculations. However, our complete and explicit computations of Feynman diagrams in Fig.~\ref{FigG2} have invalided the above argument. It is those terms in the diagrams $(b_{1,2})$ and $(d_{1,2})$ proportional to inverse powers of $m_G$ in the numerator of the massive graviton propagator that generate the dominant power-law divergent contribution to the lepton $g-2$, which was simply ignored in Ref.~\cite{Graesser:1999yg}. Indeed, the gravitational Ward identity which was closely related to the diffeomorphism invariance, {\it i.e.}, the gauge symmetry of the massless graviton, is not expected to be applied to the massive spin-2 field which does not possess any gauge symmetry at all. Therefore, our result of the lepton $g-2$ induced by the massive spin-2 field $G$ obeys the conventional power counting rule, which is in contrast with the $g-2$ contribution from an interesting model in Ref.~\cite{Arkani-Hamed:2021xlp} with extra fermions. {Moreover, in the previous studies~\cite{Graesser:1999yg,Kang:2020huh,Hong:2017tel} of the massive graviton contribution to the lepton $g-2$, it was always assumed a universal coupling of $G$ to all fields in the SM. In contrast, here we concentrate on the non-universal couplings case in which $G$ couplings to the SM particles are independent of each other, along with the dependence of the final results on different couplings. Furthermore, Refs.~\cite{Graesser:1999yg,Hong:2017tel} computed the total contributions to the muon $g-2$ from the whole tower of KK graviton states in the large extra dimensional model~\cite{Arkani-Hamed:1998jmv}, the RS model~\cite{Randall:1999ee} and the clockwork gravity~\cite{Giudice:2016yja}. Especially, Ref.~\cite{Hong:2017tel}  found that the leading-order contribution in the small extra-dimension curvature limit was universal, while the subleading contributions could reflect the geometry of extra dimensions. However, in the present work, we only consider a single massive graviton contribution to $\Delta a_\mu$ in Eq.~(\ref{G2total}) by assuming that it dominates the anomaly. If we change our viewpoint by assuming that other higher KK excitations also give rise to similar effects, then we also need to sum them together. Given that Eq.~(\ref{G2total}) only provides the leading-order contribution in the expansion in terms of $\Lambda^2$, we expect that it can give us the precise leading-order muon $g-2$ in the small curvature limit. As for the subleading-order contributions, they cannot be accurately calculated since the single-massive-graviton expression of $\Delta a_\mu$ at least up to subleading order in $\Lambda^2$ is required, which is very complicated and is out of the scope of the present paper.  }

\section{Theoretical Constraints from Perturbativity}\label{TheoreticalConstraint}
In the previous discussion, we have shown that the introduction of a massive spin-2 field $G$ to the SM can lead to new contributions to the anomalous lepton magnetic moment $\Delta a^G_\ell$, which might potentially solve the long-standing $(g-2)_{e\,,\mu}$ anomalies. However, the theory still suffers from the theoretical constraints by requiring the validity of the perturbation expansion. In this section, we shall consider this perturbativity bound on our model.

\begin{figure}[!htb]
	\centering
	\hspace{-5mm}
	\includegraphics[width=0.48\linewidth]{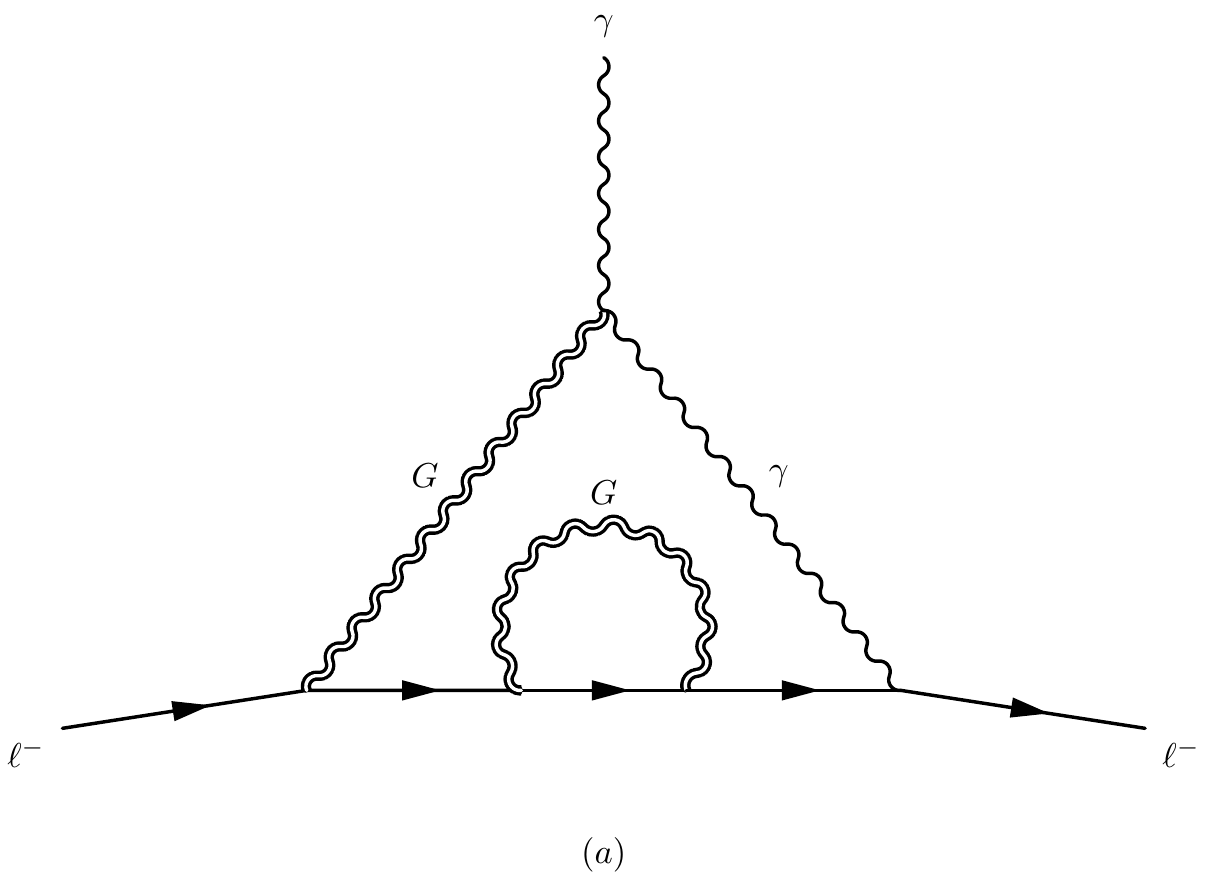}
	\includegraphics[width=0.48\linewidth]{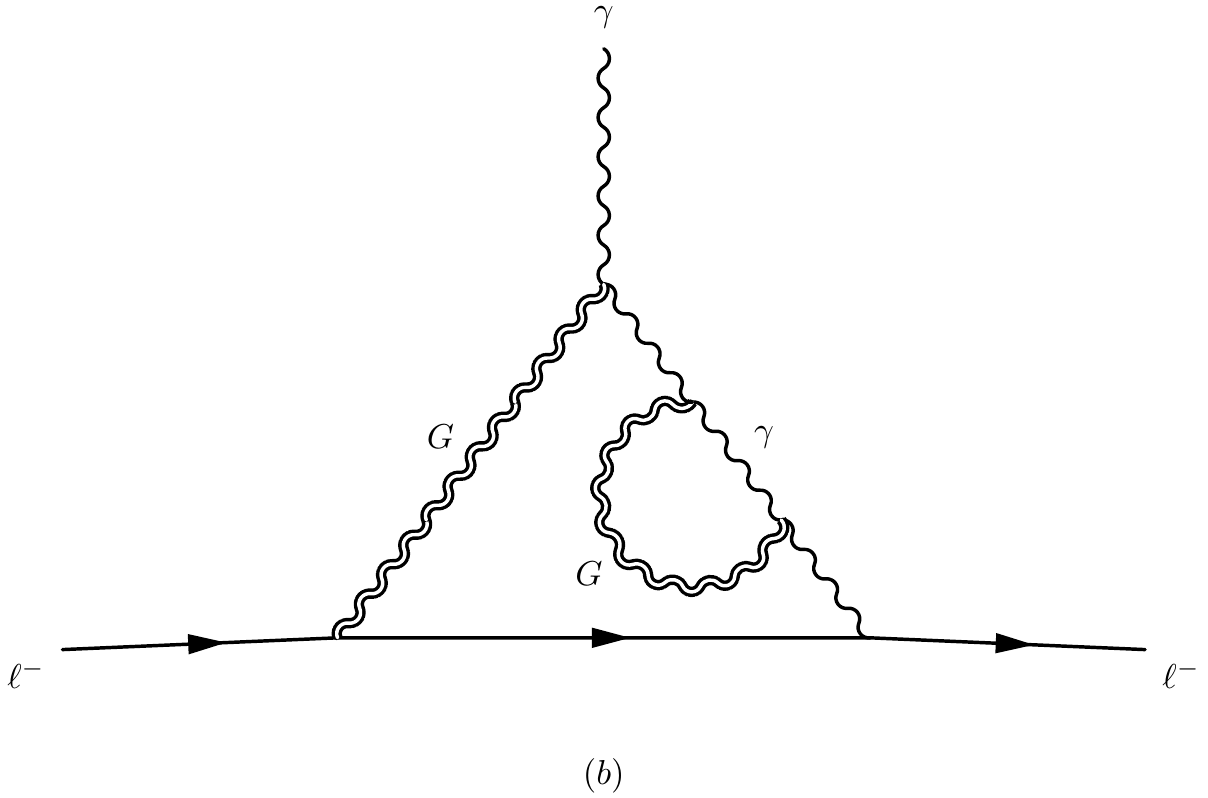}
	\caption{Two-loop Feynman diagrams that would contribute to the lepton $g-2$.}\label{Fig2Loop}
\end{figure}

Note that we have assumed implicitly that the leading-order contribution to $\Delta a_{\ell}$ comes from the one-loop Feynman diagrams induced by the massive graviton. Such a perturbativity requirement implies that the loop expansion should be valid, {\it i.e.}, the lower-loop contributions need to be larger than the higher-loop ones. For example, let us consider one particular two-loop Feynman diagram on the left panel of Fig.~\ref{Fig2Loop}, in which we attach an additional massive graviton line onto the internal lepton of the diagram $(b_1)$. 
According to the general na$\ddot{i}$ve power counting rule, we can estimate the leading-order size of $\Delta a_\mu$ as follows
\begin{eqnarray}
    \Delta_\ell^{(2)} \sim \frac{|c_\ell^3 c_\gamma|}{(16\pi^2)^2} \left(\frac{m_\ell}{\Lambda}\right)^2 \left(\frac{\Lambda}{m_G}\right)^8\,.
\end{eqnarray}
In contrast, the corresponding one-loop diagram $(b_1)$ can be order-of-magnitude estimated as follows
\begin{eqnarray}
    \Delta a_\ell^{(b_1)} \sim \frac{|c_\ell c_\gamma|}{16\pi^2} \left(\frac{m_\ell}{\Lambda}\right)^2 \left(\frac{\Lambda}{m_G}\right)^4\,.
\end{eqnarray}
Now the requirement of the perturbativity indicates that the 1-loop contribution dominates over the 2-loop one, which gives the following constraint
\begin{eqnarray}
    \frac{|c_\ell c_\gamma|}{16\pi^2} \left(\frac{m_\ell}{\Lambda}\right)^2 \left(\frac{\Lambda}{m_G}\right)^4 > \frac{|c_\ell^3 c_\gamma|}{(16\pi^2)^2} \left(\frac{m_\ell}{\Lambda}\right)^2 \left(\frac{\Lambda}{m_G}\right)^8\,,
\end{eqnarray}
which leads to 
\begin{eqnarray}\label{cmuConst}
    |c_\ell| < 4\pi \left(\frac{m_G}{\Lambda}\right)^2\,.
\end{eqnarray}
Moreover, if we apply almost the same argument to the right two-loop Feynman diagram in Fig.~\ref{Fig2Loop}, the following similar constraint to the photon-massive-gravity coupling $c_\gamma$ can be obtained
\begin{eqnarray}\label{cgConst}
|c_\gamma| < 4\pi \left(\frac{m_G}{\Lambda}\right)^2\,.
\end{eqnarray}
Note that the perturbativity constraints in Eqs.~(\ref{cmuConst}) and (\ref{cgConst}) are natural generalizations of that for a dimensionless coupling constant $g$ with its bound as $|g|<4\pi$~\cite{Nebot:2007bc}. 

\section{Numerical Studies}\label{SecResult}
Given the one-loop analytic expression of the spin-2 particle contribution to the lepton $(g-2)$ in Eq.~(\ref{G2total}) and the constraints from perturbativity presented in Sec.~\ref{TheoreticalConstraint},  we now explore the viable parameter space to explain the $(g-2)_{e,\,\mu}$ anomalies. In our study, the latest measurement of $\Delta a_\mu$ by the Muon Collaboration at Fermilab in Eq.~(\ref{ExpG2}) is taken in its $2\sigma$ allowed region. On the other hand, there are currently two incompatible theoretical predictions of the electron anomalous magnetic moment $\Delta a^{\rm LKB}_e$ at the Laboratoire Kastler Brossel (LKB)~\cite{Morel:2020dww} and $\Delta a^{\rm B}_e$ at Berkeley~\cite{Parker:2018vye}, both of which deviate the SM value substantially. Note that the difference are caused by their respective measurements of the fine structure constant $\alpha$. Thus, in what follows, we shall consider the experimental results of $\Delta a^{\rm LKB}_e$ and $\Delta a^{\rm B}_e$ separately.   
\begin{figure}[!htb]
	\centering
	\hspace{-5mm}
	\includegraphics[width=0.5\linewidth]{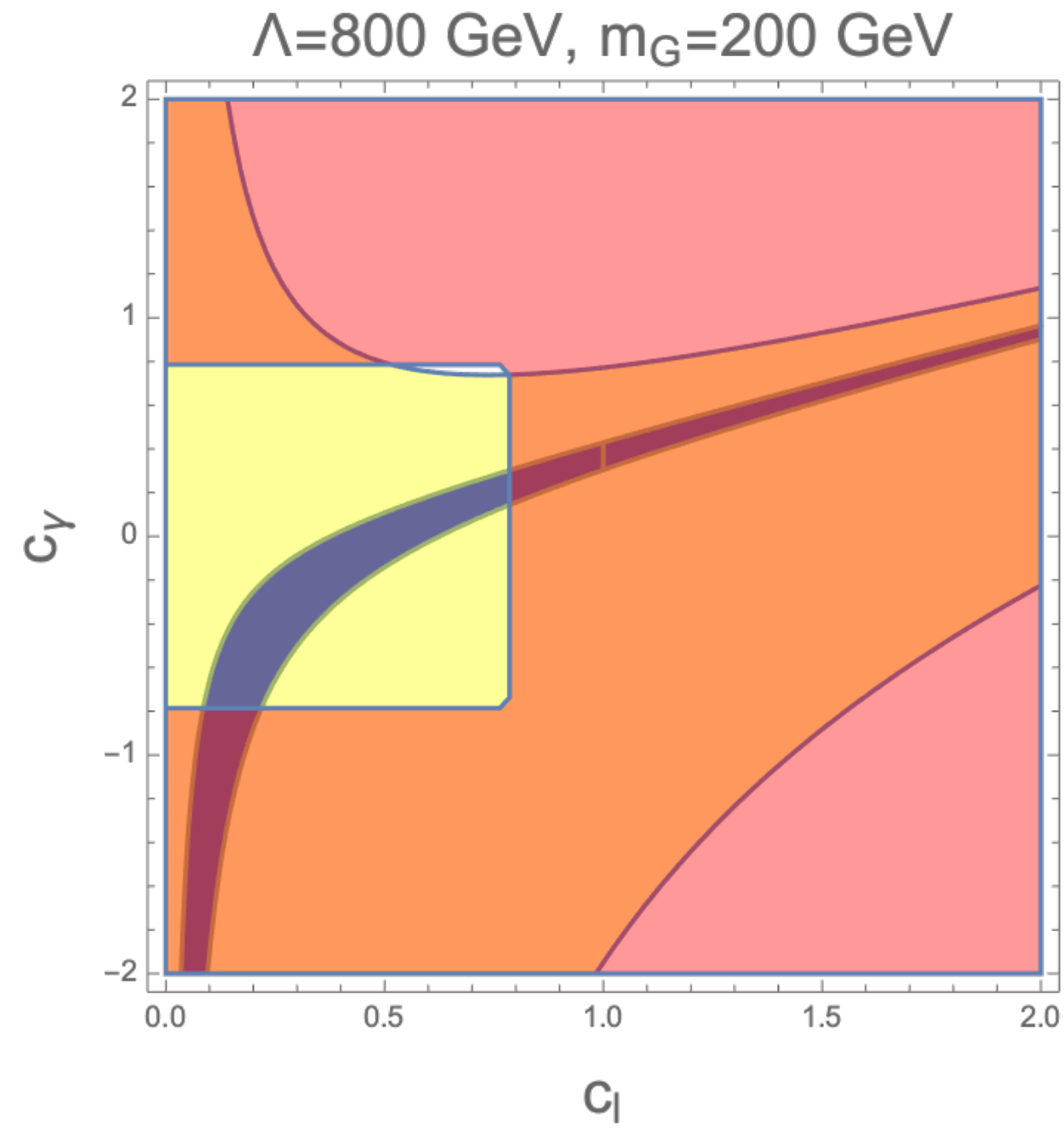}
	\includegraphics[width=0.5\linewidth]{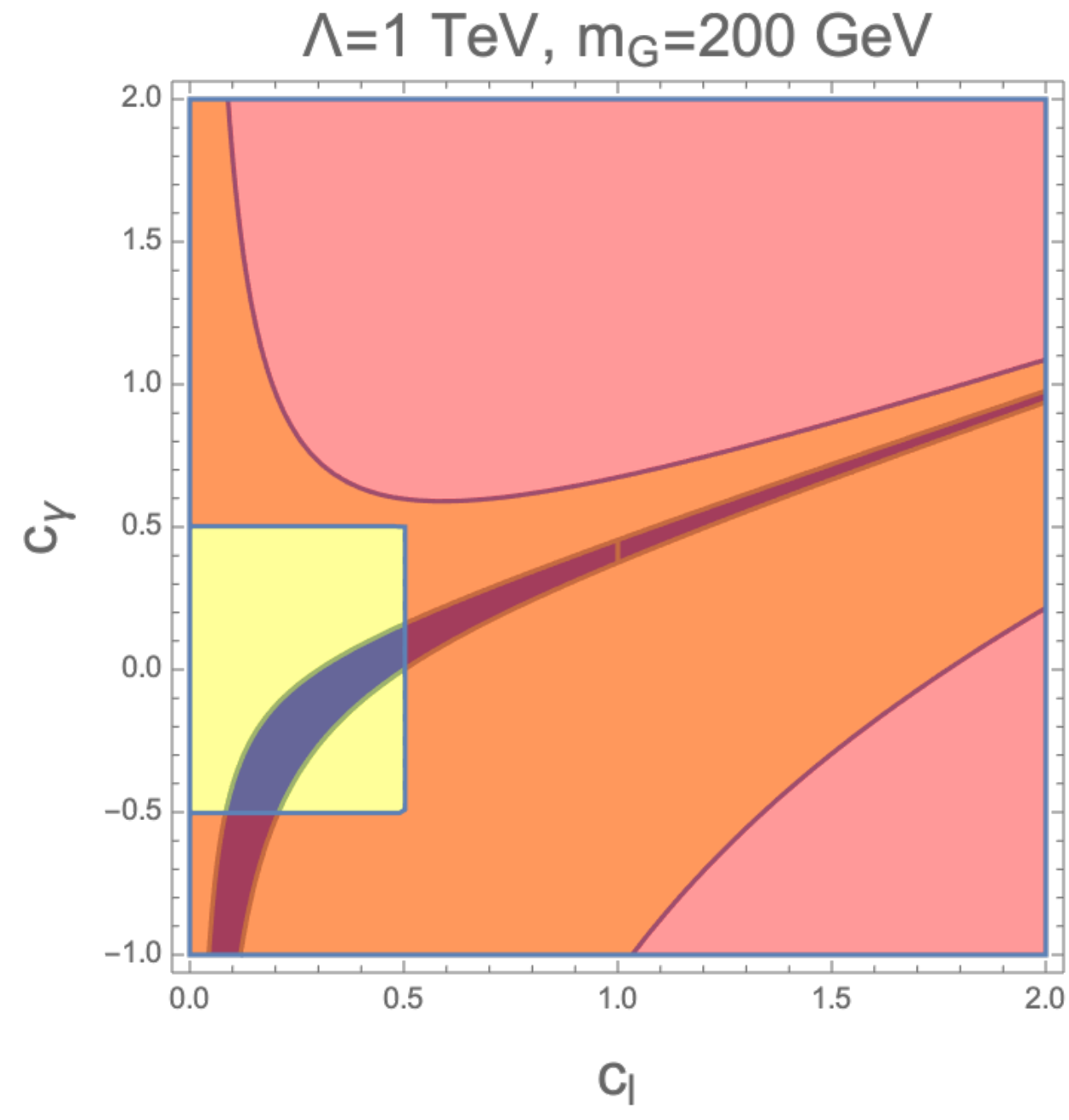}	
	\caption{The parameter space in the $c_\ell$-$c_\gamma$ plane for the massive graviton mass fixed at $m_G=200$~GeV and the cutoff scale at $\Lambda = 800$~GeV (left panel) and $1$~TeV (right panel). The blue and yellow shaded regions show the parameter space that can explain the $\Delta a_\mu$ and $\Delta a_e^{\rm LKB}$ anomalies in $2\sigma$ range, while the areas colored in red are excluded by the theoretical perturbativity constraints, respectively.} \label{M200}
\end{figure}
\begin{figure}[!htb]
	\centering
	\hspace{-5mm}
	\includegraphics[width=0.5\linewidth]{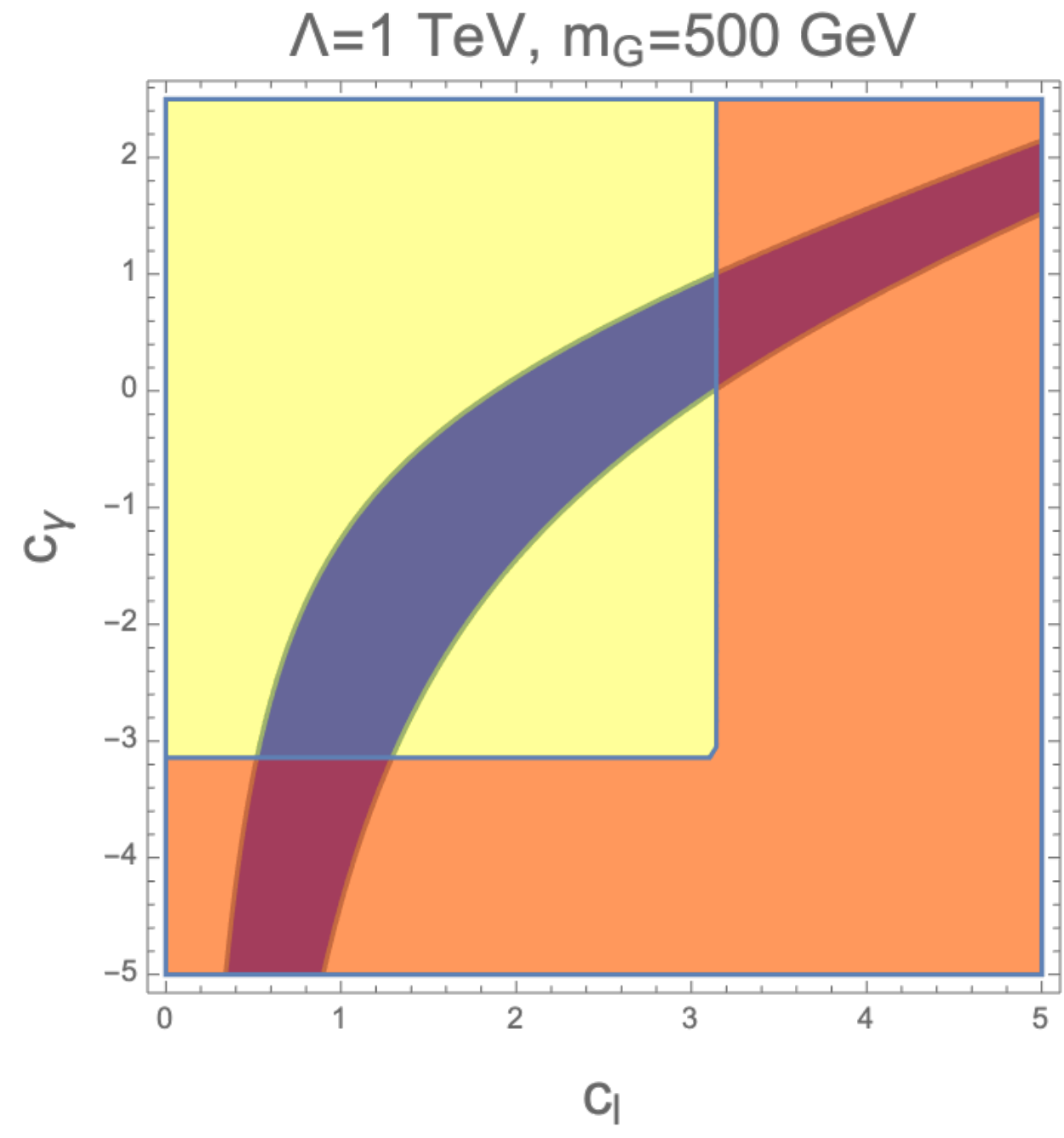}
	\includegraphics[width=0.5\linewidth]{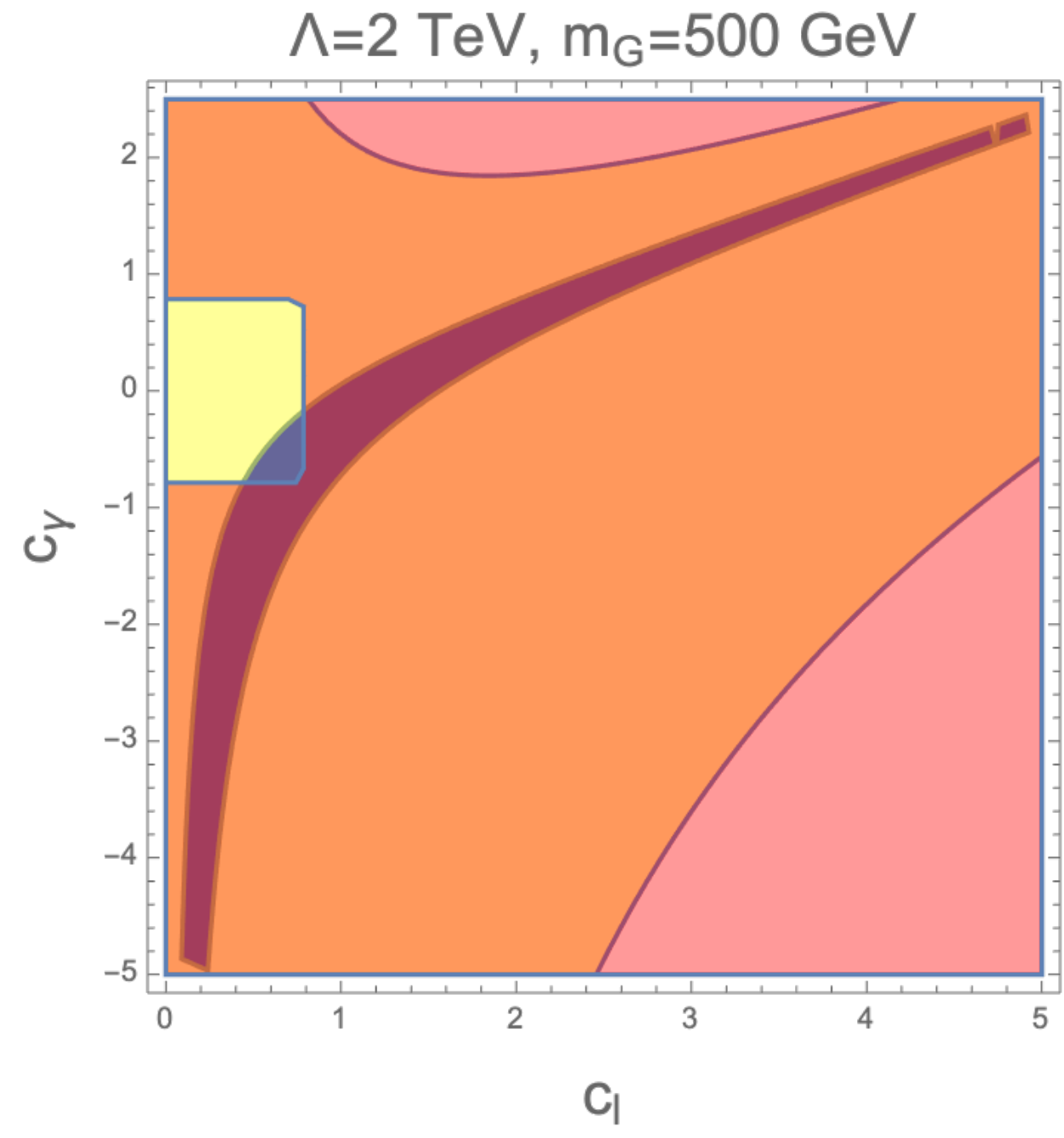}	
	\caption{The parameter space in the $c_\ell$-$c_\gamma$ plane for the massive graviton mass fixed at $m_G=500$~GeV and the cutoff scale at $\Lambda = 1$~TeV (left panel) and $2$~TeV (right panel). The color coding is the same as that in Fig.~\ref{M200}.} \label{M500}
\end{figure}

Figs.~\ref{M200} and \ref{M500} show the parameter space in the $c_\ell$-$c_\gamma$ plane for the massive graviton mass $m_G = 200$ and 500~GeV, where the cutoff scales are taken to be $\Lambda = 800$~GeV and $1$~TeV in the former case while $\Lambda = 1$ and 2 TeV in the latter. In all the plots the blue and yellow shaded regions represent those parameter space allowed by the Muon $(g-2)_\mu$ and LKB $(g-2)_e$ data at $2\sigma$ CL, while the red shaded regions are excluded by the perturbativity bounds, respectively. Note that here $c_\ell = c_\mu$ and $c_e$ when explaining the muon and electron anomalous magnetic moments, respectively. It is seen from these plots that, although the perturbativity strongly constrains $c_{\gamma,\,\ell}$, there is still an ample viable parameter space in which the massive graviton induced contributions can explain both $\Delta a_\mu$ and $\Delta a_e^{\rm LKB}$ discrepancies. By comparing the two plots in Figs.~\ref{M200} and \ref{M500}, we see that the increase of the cutoff scale $\Lambda$ for a fixed $m_G$ would make the parameter space allowed by perturbativity shrink greatly. Also,  as the spin-2 particle becomes heavier with $\Lambda$ fixed, the Wilson coefficients $c_{\ell}$ and $c_\gamma$ would be pushed into larger values in order to compensate for the $m_G$ suppression, as is evident from the $\Delta a^G_{\ell}$ formula in Eq.~(\ref{G2total}).  

\begin{figure}[!htb]
	\centering
	\hspace{-5mm}
	\includegraphics[width=0.48\linewidth]{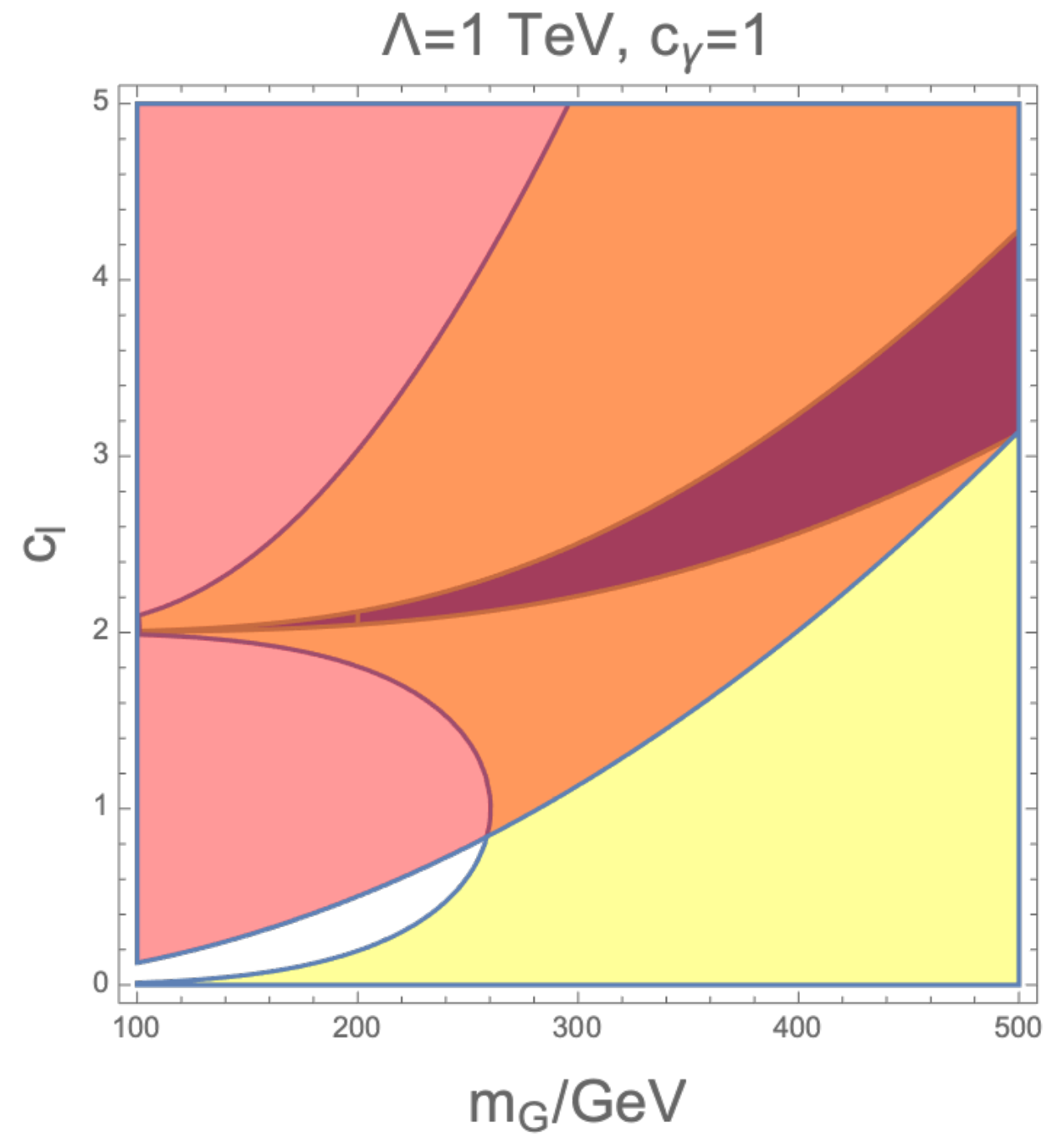}
	\includegraphics[width=0.48\linewidth]{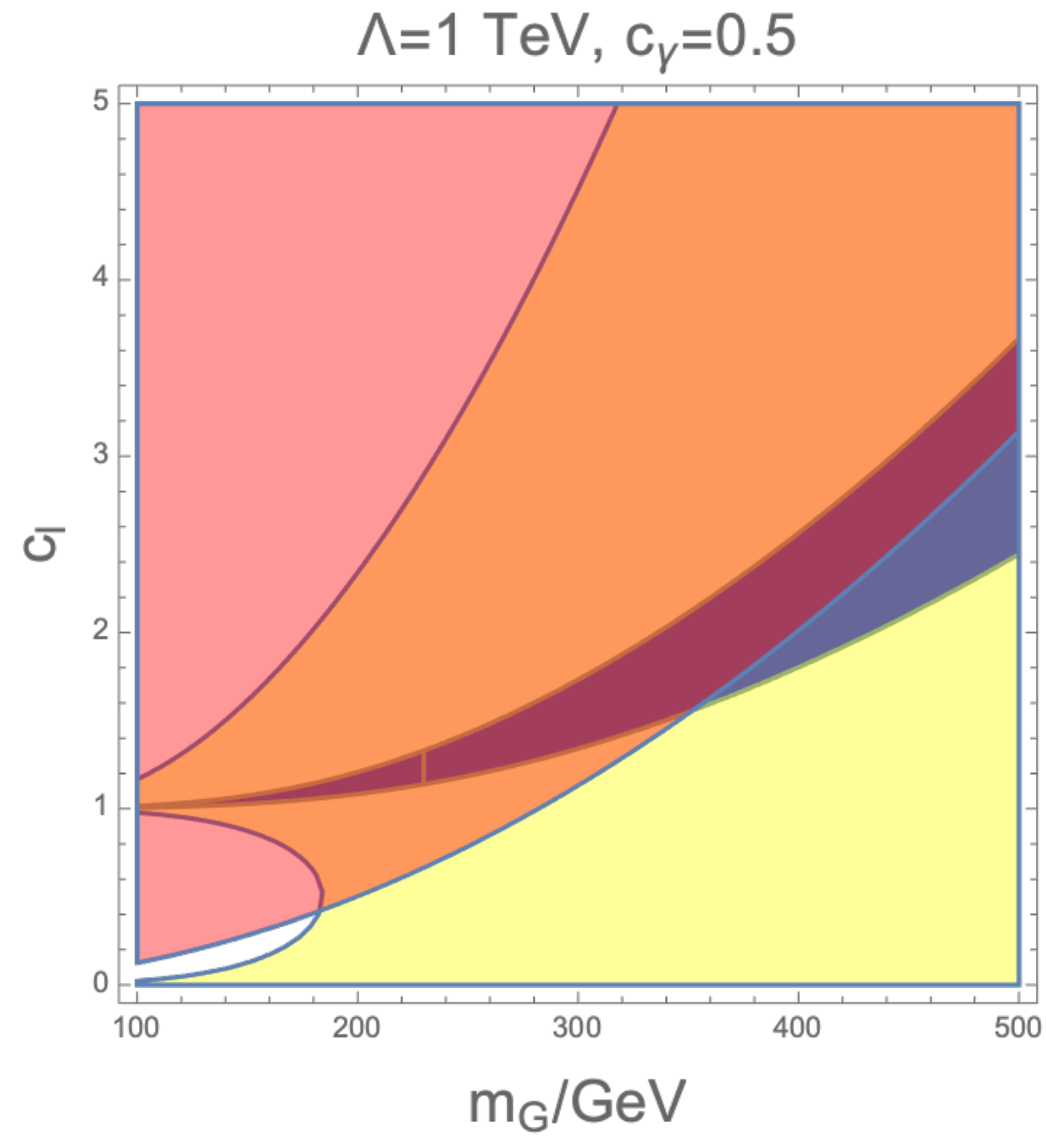}	
	\includegraphics[width=0.48\linewidth]{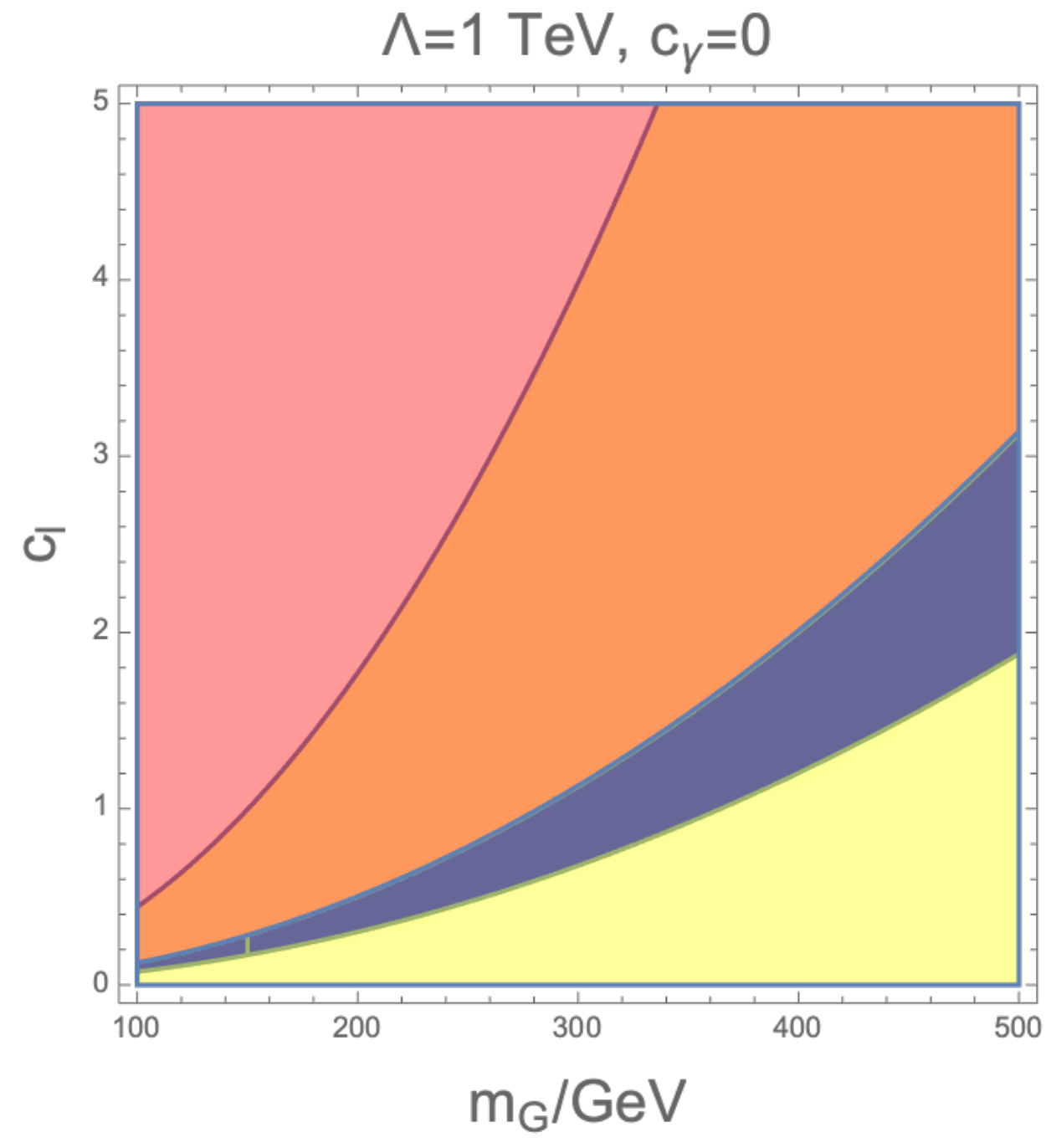}	
	\includegraphics[width=0.48\linewidth]{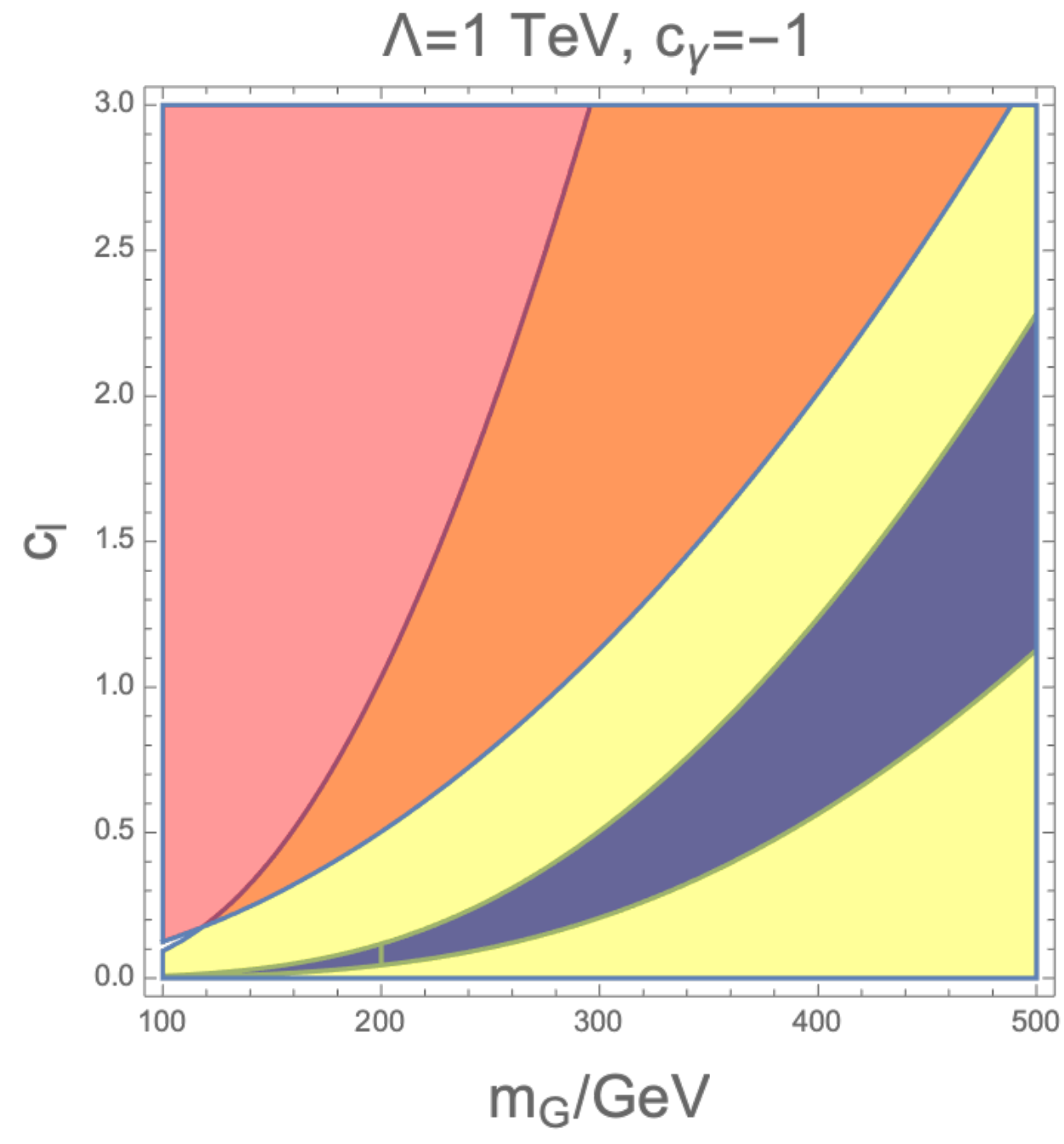}
	\caption{The parameter space in the $m_G$-$c_l$ plane with $\Lambda=1$~TeV and $c_\gamma=1$ (top-left panel), 0.5 (top-right panel), 0 (bottom-left panel), and $-1$ (bottom-right panel). The color coding is the same as that in Fig.~\ref{M200} } \label{cGplot}
\end{figure}

In Fig.~\ref{cGplot}, the relevant parameter space is also shown in the $m_G$-$c_\ell$ plane by fixing $\Lambda=1$~TeV as well as $c_\gamma = 1$, 0.5, 0, and $-1$, respectively. When $c_\gamma=1$, the blue bands that could explain the muon $g-2$ anomaly with $m_G < 500$~GeV are all excluded by the perturbativity constraints on $c_\ell$. As $c_\gamma$ decreases, more and more muon $g-2$ signal region becomes allowed by perturbativity. In particular, for the case with $c_\gamma=0$, the Barr-Zee contributions~\cite{Bjorken:1977vt,Barr:1990vd}  displayed as $(b_{1,2})$ in Fig.~\ref{FigG2} are effectively turned off, while the term induced solely by the lepton-$G$ couplings $c_\ell$ in Eq.~(\ref{G2total}) can interpret the anomalies in $\Delta a_\mu$ and $\Delta a_e^{\rm LBK}$ without disturbing any theoretical validity. {Moreover, as shown in Figs.~\ref{M200}, \ref{M500} and \ref{cGplot}, the current measurements by the Muon $g-2$ and LKB experiments still allow us to take a lepton universal coupling $c_\ell = c_e = c_\mu$ in our spin-2 particle model. However, as illustrated in the next section, when considering the collider constraints, such a lepton flavor universal case has already been disfavored by the existing data.}

\begin{figure}[!htb]
	\centering
	\hspace{-5mm}
	\includegraphics[width=0.5\linewidth]{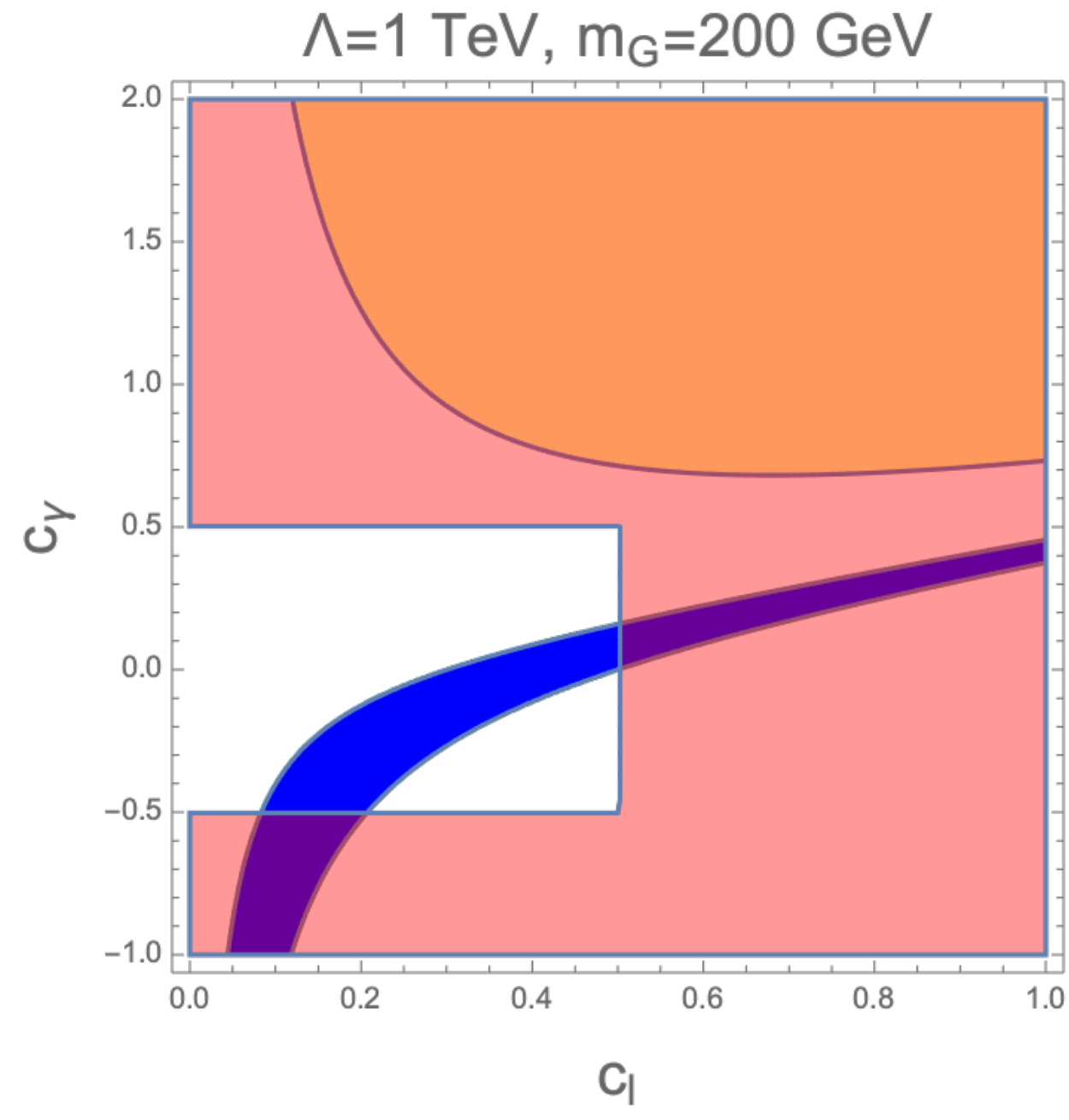}
	\includegraphics[width=0.5\linewidth]{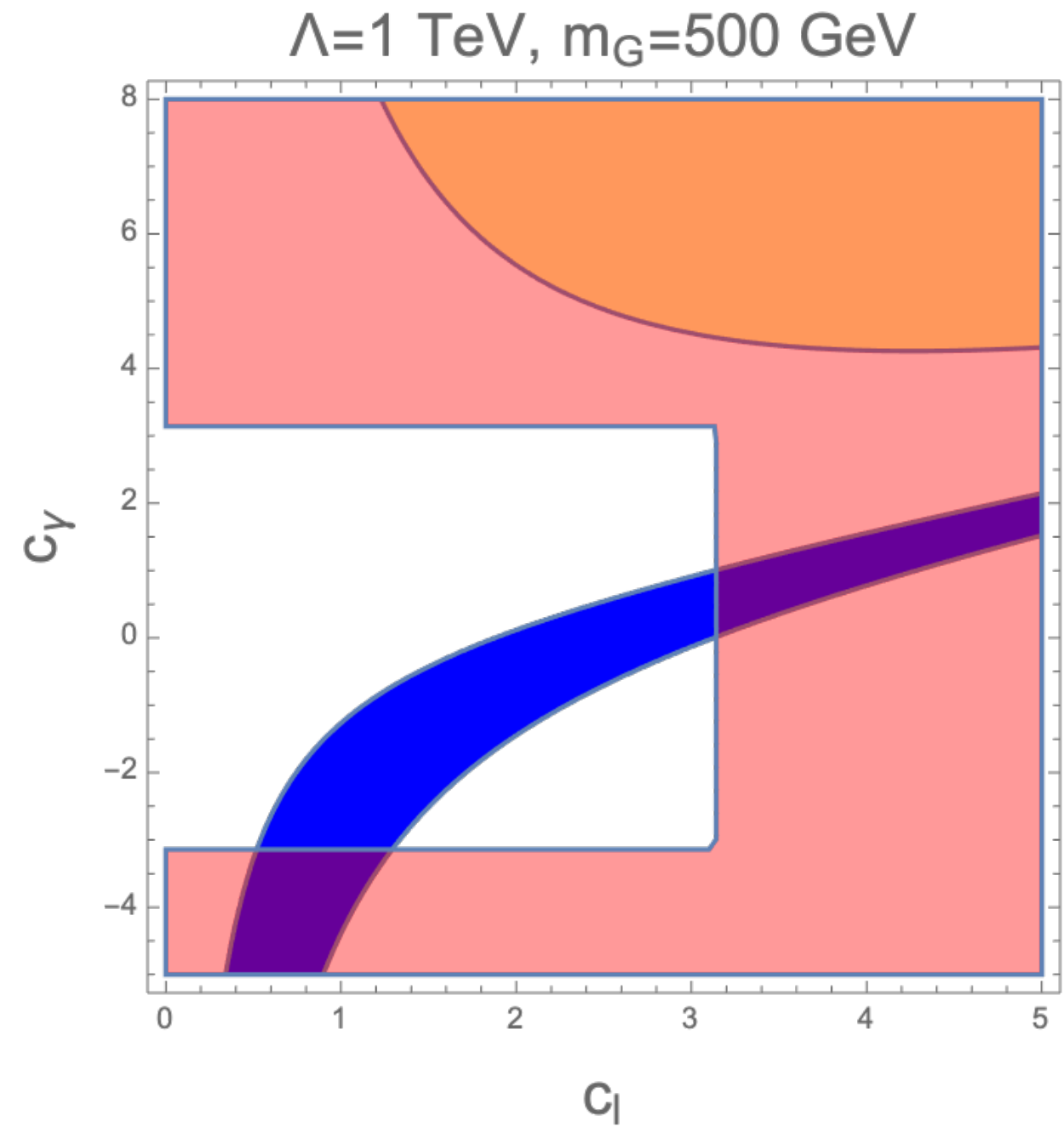}	
	\caption{The parameter space in the $c_\ell$-$c_\gamma$ plane for the cutoff scale fixed at $\Lambda = 1$~TeV and the massive graviton mass at $m_G=200$~GeV (left panel) and $500$~GeV (right panel). The color coding is the same as that in Fig.~\ref{M200}, except that the yellow region now represents the parameter space predicted by the Berkeley data of $\Delta a_e^{\rm B}$.} \label{plotBerkeley}
\end{figure}
Finally, we turn to the spin-2 particle interpretation of the $\Delta a_\mu$ anomaly and the Berkeley measurement on the electron anomalous magnetic moment $\Delta a_e^{\rm B}$, with the numerical results given in Fig.~\ref{plotBerkeley}. Note that the Berkeley measurement of the fine structure constant prefers a negative $\Delta a_e^{\rm B}$, which is in stark contrast with its positive muon counterpart $\Delta a_\mu$. Thus, it is rather difficult in explaining the $\Delta a_e^{\rm B}$ and $\Delta a_\mu$ anomalies simultaneously. In the present spin-2 particle model, the Feynman diagrams $(d_{1,2})$ always give positive contributions to the lepton anomalous magnetic moments, while the sign of $\Delta a_\ell$ from the Barr-Zee diagrams $(b_{1,2})$ depends on that of the combination $c_{\ell} c_\gamma$. Hence, if the Barr-Zee diagrams dominate the contribution to $\Delta a_e$ and $c_e c_\gamma>0$, then it provides us a nice explanation on the opposite sign between $\Delta a_\mu$ and $\Delta a_e^{\rm B}$. In this case, it usually requires a large value of $|c_\gamma|$, which has, unfortunately, been strongly disfavored by the perturbativity constraints as evident from Fig.~\ref{plotBerkeley}. Therefore, it seems that the current simple spin-2 particle explanation of the muon $g-2$ result cannot offer a viable simultaneous solution to the Berkeley anomaly on $(g-2)_e$. 

\section{Collider Constraints}\label{Collider}
Note that the explanation of the $(g-2)_{e,\,\mu}$ anomalies requires a relatively light spin-2 particle with its mass around several hundred GeV and substantially large couplings to charged leptons and photons, which indicates that this massive graviton would have considerable decay rates to the di-lepton and di-photon final states. Thus, the present massive graviton scenario is well suitable to be tested by collider experiments, such as LHC and LEP-II. In fact, there have already been many searches at the LHC for the spin-2 resonance in the dilepton and diphoton channels at both ATLAS~\cite{ATLAS:2019erb,Wang:2021rvc} and CMS~\cite{CMS:2018ipm,CMS:2018dqv,Radburn-Smith:2018wfo}. For the original RS model with a universal coupling to all SM particles, the lower bounds on the cutoff scale from the $\ell\ell$ and $\gamma\gamma$ channels have been $\Lambda/c_{\rm SM} \sim {\cal O}(100~{\rm TeV})$~\cite{Kraml:2017atm} for the massive graviton mass below 1~TeV, which has excluded the possibility to explain the lepton $g-2$ anomalies in terms of $G$. However, in the generalized RS models with non-universal SM particle couplings, the above conclusion does not apply any more. Note that the current experimental limits from LHC are only placed on $\sigma (pp\to G) \times {\cal B}(G\to \ell\ell \mbox{ or } \gamma\gamma)$ with $\sigma (pp\to G)$ and ${\cal B}$ denoting the production cross section and decay branching fractions of the spin-2 particle. If the $G$ production rate or its decay branching fractions to $\ell\ell$ and $\gamma\gamma$ are suppressed, then the above constraints on $G$ can be relaxed. For example,  in the model of Refs.~\cite{Geng:2016xin,Geng:2018hpq}, the unconventional power counting rule predicts the dominant massive graviton decay channel is $t\bar{t}$, and the branching fraction of $G$ decaying to $\gamma\gamma$ is given by ${\cal B}(G\to \gamma\gamma) \sim 10^{-4}$, which leads to a much lower available cutoff scale $\Lambda/c_\gamma > {\cal O}(200~{\rm GeV})$ from the extension of the diphoton upper bound in Fig.~6 of Ref.~\cite{Geng:2018hpq} to the low $m_G$ region. Such a small cutoff scale is exactly what is needed to explain the $(g-2)_{\mu,\,e}$ anomalies. {More recently, both ATLAS and CMS have updated their resonance searches in the channels such as $t\bar{t}$~\cite{ATLAS:2018rvc,CMS:2018rkg}, dijet~\cite{ATLAS:2019fgd,CMS:2019gwf}, diboson~\cite{ATLAS:2019nat,CMS:2019qem}, diphoton~\cite{Wang:2021rvc,CMS:2018dqv} and dilepton~\cite{ATLAS:2019erb,CMS:2018ipm}, which have made the lower bounds on the cutoff scale in Ref.~\cite{Geng:2018hpq} somewhat outdated. Nevertheless, we still expect that there is still much room for the cutoff $\Lambda \lesssim 1$~TeV available to accommodate the spin-2 particle explanation of the muon $g-2$. }

{The present spin-2 particle model can also be tested by LEP-II~\cite{ALEPH:2013dgf,ALEPH:2006bhb,DELPHI:2005wxt}, which was an electron-positron collider with its center-of-mass energy running from the $Z$ pole up to 209~GeV. In particular, the LEP-II experiments have investigated the quantum gravity model by measuring the total and differential cross sections in the $\gamma\gamma$, $e^+ e^-$ and $\mu^+ \mu^-$ channels~\cite{ALEPH:2013dgf}, which are the most relevant to our present study. Following the conventions in Refs.~\cite{ALEPH:2013dgf}, the constraint for each channel is placed on the parameter $\epsilon \equiv \lambda/M_s^4$, where $\lambda$ is a dimensionless coefficient of ${\cal O}(1)$ and $M_s$ is the gravitational mass scale. For references, we list the corresponding constraints for different channels as follows. For the diphoton final state, the limit is $M^\gamma_s > 868 (1108)$~GeV for $\lambda=\pm 1$~\cite{ALEPH:2013dgf}, where $\lambda = \pm 1$ correspond to the cases of positive and negative interferences with the SM amplitudes, respectively, and the superscript on $M_s$ denotes the channel. For the $e^+e^-$ channel, $M^e_s > 1.09 (1.25)$~TeV when $\lambda = \pm 1$~\cite{ALEPH:2013dgf}, while for $\mu^+\mu^-$, $M^\mu_s > 0.695(0.793)$~TeV for $\lambda=\pm 1$~\cite{DELPHI:2005wxt}. All the upper limits are given at the 95\% confidence level. In the present model, we can approximately express $\epsilon$ in terms of model parameters as follows
\begin{eqnarray}
		\epsilon^{\gamma,e,\mu} \equiv \frac{\lambda}{(M_s^{\gamma,e,\mu})^4} \approx \frac{c_e c_{\gamma,e,\mu}}{\Lambda^2 m_G^2}\,,
\end{eqnarray}
where the massive graviton mass $m_G$ is assumed to be larger than $\sim 200$~GeV so that the internal $G$ propagator can be contracted into a point-like contact interaction. Then we can map the LEP-II experimental constraints of the extra-dimensional models onto our parameter space of interest, which are shown in Fig.~\ref{plotLEP}. 
\begin{figure}[!htb]
		\centering
		\hspace{-5mm}
		\includegraphics[width=0.5\linewidth]{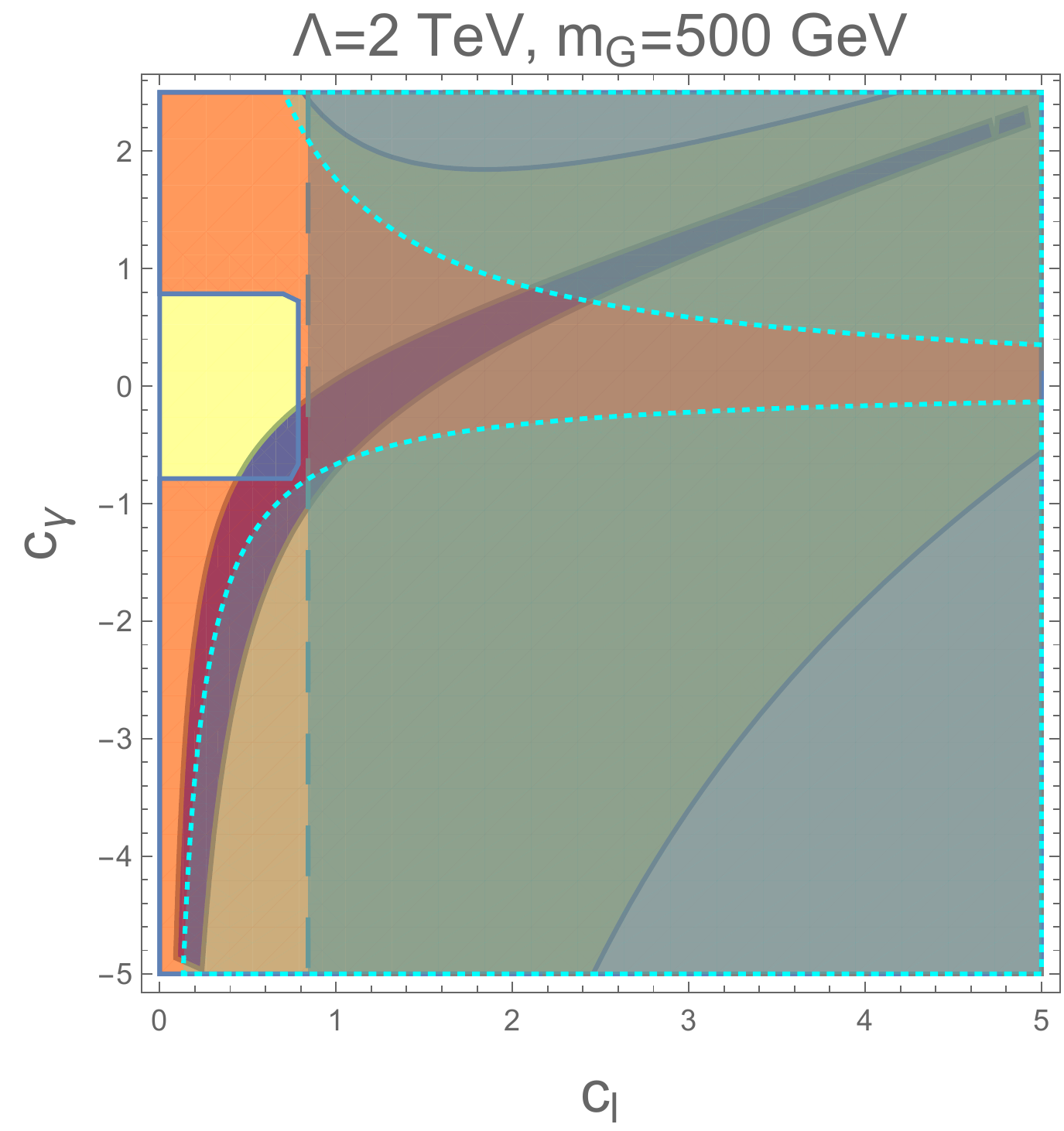}
		\includegraphics[width=0.5\linewidth]{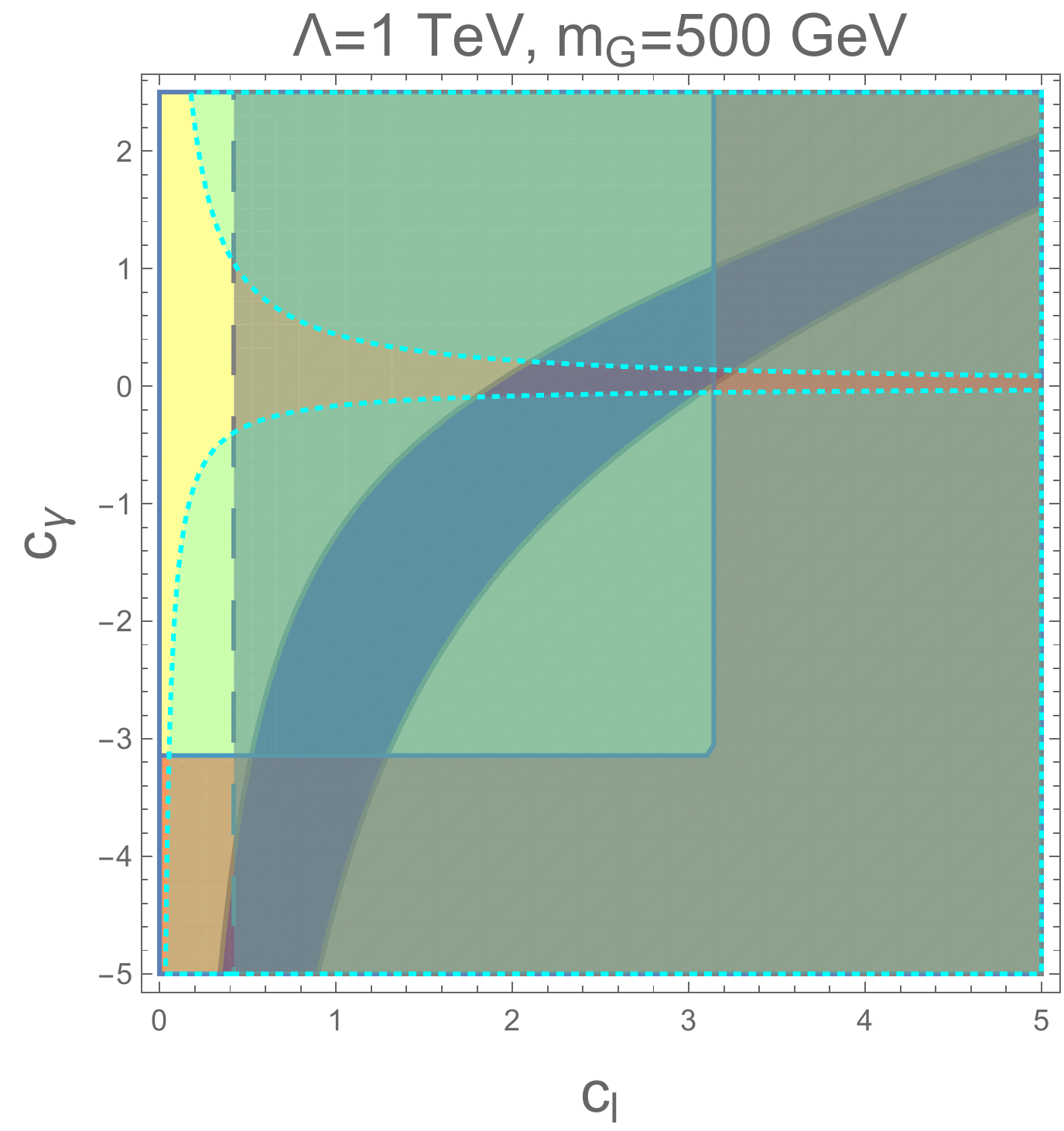}	
		\caption{The parameter space in the $c_\ell$-$c_\gamma$ plane for the graviton mass fixed at $m_G = 500$~GeV and the cutoff scale at $\Lambda=2$~TeV (left panel) and $1$~TeV (right panel) when considering the LEP-II constraints. The gray and cyan areas are excluded by the LEP-II lower bounds on the gravitational mass scale $M_s$ from the $e^+ e^-$ and $\gamma\gamma$ final states with gray dashed and cyan dotted curves denoting their respective boundaries. Other color codings are the same as those in Fig.~\ref{M200}.} \label{plotLEP}
\end{figure}
Note that here we have only used the lower bounds on $M_s$ from the $\gamma\gamma$ and $e^+e^-$ channels since they are more restrictive than the one from $\mu^+\mu^-$.  Even though we can still find very limited parameter regions as in the left panel of Fig.~\ref{plotLEP} such that the muon $g-2$ anomaly and the LKB $(g-2)_e$ data can be explained with a universal coupling $c_\ell = c_e = c_\mu$, most parameter space for this kind of models has been excluded by the LEP-II data, especially for the cutoff scale $\Lambda \lesssim 1$~TeV as illustrated on the right panel. Therefore, the massive graviton explanation of the muon $g-2$ anomaly with a lepton flavor-blind coupling $c_\ell$ is now disfavored in view of the LEP-II constraints. Here we should emphasize that the LEP-II measurements can only restrict models with $c_\mu$ of similar size as $c_e$, especially for the lepton-universal coupling case. When $|c_e| \lesssim {\cal O}(0.1)$, the LEP constraints can be easily evaded, while the model is not in conflict with the LKB $(g-2)_e$ data. The couplings $c_\mu$ and $c_\gamma$ chosen as in the blue bands in Figs.~\ref{M200}, \ref{M500} and \ref{cGplot} can still provide us a viable explanation on the muon $g-2$ anomaly.  
}

\section{Conclusions and Discussions}\label{Conclusion}
Motivated by the latest measurement on the muon anomalous magnetic moments $(g-2)_\mu$ by the Muon $g-2$ Collaboration at Fermilab~\cite{Muong-2:2021ojo} and those of electrons $(g-2)_e$ at LKB~\cite{Morel:2020dww} and Berkeley~\cite{Parker:2018vye}, we have explored the possibility to explain these anomalies in terms of the presence of a massive spin-2 particle $G$, which can be identified as the first KK excitation of the ordinary graviton in the generalized RS scenario. By calculating the associated one-loop Feynman diagrams, we have given the analytic expression of the leading-order contributions to the lepton anomalous magnetic moments induced by $G$. Note that the integrals over the loop momentum in the Feynman diagrams are all of highly power-law divergence, {\it i.e.}, of ${\cal O}(\Lambda^4)$, with $\Lambda$ representing the UV cutoff scale. Moreover, in the Barr-Zee type diagrams~\cite{Bjorken:1977vt,Barr:1990vd}, the gauge invariance involving the internal photon line should be preserved, which is another difficulty in our computation. In order to keep the gauge invariance and the power-law divergence structure, the loop regularization method~\cite{Wu:2002xa,Wu:2003dd} has been applied. In particular, we have explicitly checked the photon gauge invariance by performing our calculation of Barr-Zee diagrams in both the Feynman-'t Hooft gauge and the general gauge with the parameter $\xi$ free. {Based on our general formulas for the spin-2 particle contribution to the lepton $g-2$, we have performed phenomenological studies on the present model. We have considered the theoretical bounds from the perturbativity and the experimental constraints from LHC and LEP-II. Interestingly, we have given a new cutoff-dependent perturbativity constraint on the associated Wilson coefficients of the nonrenormalizable operators, which is a natural but nontrivial generalization of the counterpart for dimensionless renormalizable operators. As a result, we have shown that there exists a substantial amount of parameter space to accommodate the muon $g-2$ anomaly without disturbing the perturbativity and collider constraints. Note that the $(g-2)_\mu$ anomaly and the LKB $(g-2)_e$ data still allow the spin-2 particle couplings to leptons to be universal, {\it i.e.}, $c_\ell = c_e = c_\mu$, which is, however, disfavored by the existing LEP-II data. Moreover, the present simple massive graviton framework seems impossible to take into account the Berkeley's result of the electron's anomalous magnetic moment, due to the strong theoretical bounds from the perturbativity. }


Besides the constraints from the perturbativity, another criterion to determine if our perturbative calculations  remain under control is the tree-level unitarity bounds~\cite{Gell-Mann:1969cuq,Weinberg:1971fb,Lee:1977yc,Lee:1977eg}, which give extra constraints to our spin-2 particle model~\cite{Falkowski:2016glr}. However, the examination of unitarity bounds requires the careful calculation of 2-to-2 scattering amplitudes of $\ell^-\ell^+ \to \ell^-\ell^+$ and  $\gamma\gamma\to \gamma\gamma$ for various helicity assignments of external particles. However, since the detailed calculation of unitarity bounds is rather involved, we would like to discuss it in a separate work which is still under progress~\cite{Huang0}.

{Finally, we would like to mention several salient features in the flavor physics for the massive graviton model with a lepton-universal coupling, though it is not favored by the LEP-II data.} As shown in Sec.~\ref{model}, the general theory of the massive graviton admits the flavor off-diagonal terms between $G$ and charged leptons in Eqs.~(\ref{OffDiagInt}) and (\ref{OffDiagET}), which would give rise to the LFV processes, such as $\mu \to e\gamma$~\cite{MEG:2016leq}, $\mu^+ \to e^+ e^+ e^-$~\cite{SINDRUM:1987nra}, and $\mu$-$e$ conversions in nuclei~\cite{SINDRUMII:2006dvw}. Moreover, the Wilson coefficients of these effective interactions $c_{\ell^\prime \ell}$ are generically complex, so that they would also lead to CPV observables like the electron and muon EDMs~\cite{ACME:2018yjb,Muong-2:2008ebm}. Therefore, it is generally expected that such lepton flavor off-diagonal terms are stringently constrained. 
However, when the $G$-lepton couplings are flavor universal,  {\it i.e.}, $c_e=c_\mu = c_\tau$, the total energy-momentum tensor $T^{\mu\nu}_\ell$ of the charged lepton sector coupled to $G$ is the same as that defined in the SM. If the charged lepton mass matrix is diagonalized in one basis, this property would be inherited by the $G$-charged-lepton couplings with the associated energy-momentum tensor. Then the LFV processes can only be induced by the interactions with active neutrinos so that they are well-known to be highly suppressed by the tiny neutrino masses, remaining to be unobservable under the present experimental status. Furthermore, due to the self-Hermitian nature of the total charged lepton energy-momentum tensor $T^{\mu\nu}_\ell$, its universal coupling to $G$ can only be real, which automatically avoids the appearance of CPV vertices. {Unfortunately, such an interesting case is now well restricted by the LEP-II data.} 
As far as we know, there is not any other natural mechanisms to forbid the complex flavor off-diagonal couplings as well as the associated LFV and CPV effects. The detailed discussion of these flavor issues is out of the scope of the present work, and we would like to leave it for future researches. 
  

\section*{Acknowledgements}
This work is supported in part by the National Key Research and Development Program of China (Grant No. 2020YFC2201501 and No. 2021YFC2203003). DH is also supported in part by the National Natural Science Foundation of China (NSFC) under Grant No. 12005254 and the Key Research Program of Chinese Academy of Sciences under grant No. XDPB15.

\appendix

\section{Calculation Details of Lepton $(g-2)$ from the Spin-2 Particle Loop Diagrams with Loop Regularization}\label{AppG2}

In this appendix, we present the calculation details for one-loop spin-2 particle induced contributions to the lepton $g-2$ shown in Fig.~\ref{FigG2}. Note that the relevant amplitude can be parameterized as follows
\begin{eqnarray}\label{g2Def}
	i{\cal M}(\gamma \ell \to \ell) = \bar{u}(k_2) (-iQ_\ell) \left[e\gamma_\mu F_1(q^2) + \frac{ie \sigma_{\mu\nu} q^\nu}{2m_\ell} F_2(q^2)\right] u(k_1)\,,
\end{eqnarray}
where $Q_\ell = -1$ and $m_\ell$ stand for the electric charge and mass of the charged lepton $\ell$, and its magnetic moment can be obtained by taking $a_\ell = F_2(0)$. Our task now is to compute the massive spin-2 particle induced contribution to this amplitude and the associated $(g-2)_\ell$.  

\subsection{Diagram $(a)$}
Now we compute the Feynman diagram ($a$) in which the loop is obtained by inserting two $G$-$\ell$ interaction vertices, with the corresponding amplitude given by
\begin{eqnarray}
	i{\cal M}_a &=& \bar{u}(k_2) \left(-\frac{ic_\ell}{4\Lambda}\right)\left[\gamma^\rho (2k_2^\sigma + l^\sigma) + \gamma^\sigma (2k_2^\rho + l^\rho)-2\eta^{\rho \sigma} (2\slashed{k}_2+\slashed{l}-2m_\ell) \right]\nonumber\\
	&& \frac{i}{\slashed{l}+\slashed{k}_2 -m_\ell} (-ie Q_\ell \gamma^\mu) \frac{i}{\slashed{l}+\slashed{k}_1 -m_\ell} \left( -\frac{ic_\ell}{4\Lambda} \right) \nonumber\\
	&& [\gamma^\lambda (2k_1^\nu + l^\nu) + \gamma^\nu(2k_1^\lambda+l^\lambda) - 2\eta^{\lambda\nu}(2\slashed{k}_1 + \slashed{l}-2m_\ell)]u(k_1) \frac{i}{2}\frac{B_{\rho\sigma,\lambda\nu}(l)}{l^2 -m_G^2} \,, 
\end{eqnarray}
where the massive spin-2 particle propagator is defined with the following factor 
\begin{eqnarray}\label{PropG}
	B_{\mu\nu\, \rho\sigma} (k) &=& \left(\eta_{\mu\rho} - \frac{k_\mu k_\rho}{m_G^2}\right) \left(\eta_{\nu\sigma} - \frac{k_\nu k_\sigma}{m_G^2}\right) + \left(\eta_{\mu\sigma} - \frac{k_\mu k_\sigma}{m_G^2}\right) \left(\eta_{\nu\rho} - \frac{k_\nu k_\rho}{m_G^2}\right) \nonumber\\
	&& -\frac{2}{3} \left(\eta_{\mu\nu} - \frac{k_\mu k_\nu}{m_G^2}\right) \left(\eta_{\rho\sigma} - \frac{k_\rho k_\sigma}{m_G^2}\right) \,.
\end{eqnarray}
For the denominator we can complete the square of the loop momentum $l$ as follows
\begin{eqnarray}
	&&\frac{1}{(l^2-m_G^2)[(l+k_2)^2 -m_\ell^2][(l+k_1)^2 -m_\ell^2]} \nonumber\\
	&=& \Gamma(3) \int dx dy \frac{1}{[(l+xk_1+yk_2)^2 -\Delta_a]^3} \,, 
\end{eqnarray}
where 
\begin{equation}
	\Delta_a = (x+y)^2 m_\ell^2 -xy q^2 + (1-x-y)m_G^2 \,.
\end{equation}
Therefore, we can shift the loop momentum as $l \to l-x k_1 -y k_2$, which gives the following numerator of the loop integral
\begin{eqnarray}
 && \bar{u}(k_2) \left\{ \gamma^\rho [l -x k_1 +(2-y)k_2]^\sigma + \gamma^\sigma [l-x k_1 +(2-y)k_2]^\rho - 2 \eta^{\rho\sigma} [\slashed{l} - x\slashed{k}_1 + (2-y)\slashed{k}_2 -2 m_\ell] \right\} \nonumber\\
	&& [\slashed{l} - x\slashed{k}_1 + (1-y)\slashed{k}_2 + m_\ell] \gamma^\mu [\slashed{l}+(1-x)\slashed{k}_1 -y \slashed{k}_2 + m_\ell] \nonumber\\
	&& \left\{ \gamma^\lambda [l + (2-x)k_1 -yk_2]^\nu + \gamma^\nu [l + (2-x) k_1 -y k_2]^\lambda - 2 \eta^{\lambda \nu} [\slashed{l} + (2-x)\slashed{k}_1 -y \slashed{k}_2 - 2m_\ell]\right\} u (k_1) \nonumber\\
	&& B_{\rho \sigma, \lambda \nu} (l-xk_1 -yk_2)\,,
\end{eqnarray}
where we have omitted the factors proportional to the coupling constants and the electric charge. 
In our work, we only focus on the leading-order contribution to the muon anomalous magnetic moment $(g-2)_\mu$. Note that the above loop integral is highly divergent, so that it should be dominated by terms of the largest divergence degree. Note also that the terms of interest should be proportional to $\bar{u}(k_2) (i\sigma^{\mu\nu} q_\nu) u(k_1)$ which flips the lepton chirality in the amplitude. However, the only operator that can achieve this in the present model is the lepton mass term so that the final expression should be proportional to $m_\mu$. Therefore, we need to pick terms which are divergent with two powers less than the top divergence of ${\cal O}(\Lambda^8)$, {\it i.e.}, the terms with ${\cal O}(l^6)$ in the numerator. With {\sf FeynCalc}, we can obtain the following relevant terms:
\begin{eqnarray}
	\frac{64 l^4}{ 2 m_G^4} \bar{u}(k_2) \left\{ -(k_1\cdot l) \slashed{k}_2 \slashed{l} \gamma^\mu - (k_2\cdot l) \gamma^\mu \slashed{l} \slashed{k}_1 + m_\mu (k_1\cdot l)  \slashed{l} \gamma^\mu  + m_\ell (k_2\cdot l) \gamma^\mu \slashed{l}   \right\} u(k_1)=0 \,,\,\,\,
\end{eqnarray}
where we have used the equations of motion of leptons $\bar{u}(k_2) \slashed{k}_2 = m_\ell \bar{u}(k_2)$ and $\slashed{k}_1 u(k_1) = m_\ell u(k_1)$. Note that here we have not listed all terms at this order but only those which might potentially generate the desired dipole operator.  Therefore, at the order of quartic divergence, the diagram $(a)$ does not give any contribution to $\Delta a_\ell$.  


\subsection{Diagram $(b_1)$ in the Feynman-'t Hooft Gauge}
Let us turn to the contributions from the Barr-Zee diagrams~\cite{Bjorken:1977vt,Barr:1990vd}, $(b_1)$ and $(b_2)$. Since they are symmetric under the swapping of two vertices along the fermion line, their contributions are expected to be equal. This subsection is devoted to computing the diagram $(b_1)$. The associated amplitude is given as follows:
\begin{eqnarray}\label{ExpB1}
	i{\cal M}_{(b_1)} &=& \bar{u}(k_2) (-ieQ_\ell \gamma^\nu) \frac{i}{\slashed{l}-m_\ell} \left(-\frac{ic_\ell}{4\Lambda}\right) \left[\gamma^\rho (l+k_1)^\sigma + \gamma^\sigma (l+k_1)^\rho -2\eta^{\rho \sigma} (\slashed{l}+\slashed{k}_1 -2m_\ell)\right] u(k_1) \nonumber\\
	&& \frac{-iA_{\nu\kappa}(l-k_2)}{(l-k_2)^2} \left(-\frac{ic_\gamma}{\Lambda}\right) \left[q\cdot (l-k_2) C^{\alpha \beta, \mu\kappa} + D^{\alpha \beta, \mu\kappa}(q, l-k_2) + \xi^{-1} E^{\alpha\beta, \mu \kappa}(q,l-k_2)\right] \nonumber\\
	&& \frac{i}{2} \frac{B_{\alpha\beta, \rho\sigma}(l-k_1)}{(l-k_1)^2-m_G^2} \nonumber\\
	&=& -\frac{c_\ell c_\gamma e Q_\ell}{8\Lambda^2} \frac{\bar{u}(k_2)[\gamma^\rho (l+k_1)^\sigma + \gamma^\sigma (l+k_1)^\rho -2 \eta^{\rho \sigma}(\slashed{l}+\slashed{k}_1 -2m_\ell)]u(k_1)}{[l^2-m_\ell^2](l-k_2)^2[(l-k_1)^2-m_G^2]} \times \nonumber\\
	&& A_{\nu\kappa}(l-k_2) \left[q\cdot (l-k_2) C^{\alpha\beta, \mu\kappa} +D^{\alpha\beta,\mu\kappa}(q, l-k_2) + \xi^{-1} E^{\alpha\beta, \mu \kappa}(q,l-k_2)\right] B_{\alpha\beta,\rho\sigma}(l-k_1)\,, \nonumber\\
\end{eqnarray} 
where 
\begin{eqnarray}\label{PropGamma}
	A_{\mu\nu} (p) = \eta_{\mu\nu} - (1-\xi) \frac{p_\mu p_\nu}{p^2}\,,
\end{eqnarray}
with $\xi$ the gauge parameter. There are two possible problems related to this Barr-Zee diagram. Firstly, the amplitude is still highly divergent, so that the leading-order contribution to the lepton $g-2$ is expected to come from the largest divergence. Thus, it is important to keep the divergence structure of the loop integral. Secondly, this expression of Eq.~(\ref{ExpB1}) explicitly depends on the gauge parameter $\xi$. However, the gauge invariance of the quantum electrodynamics requires the final result should not rely on the choice of this parameter. It is well-known that the traditional dimensional regularization, even though preserving the gauge invariance, cannot keep the divergences of positive powers all of which are distorted into the logarithmic ones. Therefore, it cannot be applied here. In the literature, one method which can retain both the gauge invariance of a general gauge theory and the divergence structure is the loop regularization~\cite{Wu:2002xa,Wu:2003dd}. In the following, we shall make use of the loop regularization to perform our calculations.
 

Note that the loop regularization guarantees the gauge invariance by demanding the consistency relations~\cite{Wu:2002xa,Wu:2003dd} among the tensor and scalar-type loop integrals defined below, rather than at the Lagrangian level like in the case of dimensional regularization. One can only check the gauge invariance of the theory by explicit calculations in different gauges. In this subsection, we shall take the Feynman-'t Hooft gauge with $\xi=1$, so that $A_{\mu\nu}(p^2) = \eta_{\mu\nu}$ in order to simplify our calculation. We leave the discussion of computation details with the general gauge in the next subsection. Moreover, when using the consistency conditions in the loop regularization in Eqs.~(\ref{IdentityLR}), we may encounter 0 in the denominator for quartically divergent integrals. One way to avoid such a problem is to relax the power index of the massive graviton's propagator as follows
\begin{eqnarray}\label{PropGN}
	\frac{i}{2} \frac{B_{\mu\nu,\, \rho \sigma}(k)}{k^2-m_G^2} \to \frac{i}{2} \frac{B_{\mu\nu,\, \rho \sigma}}{[k^2-m_G^2]^n}\,,
\end{eqnarray} 
and then to take the limit $n\to 1$ in the end of the calculation. The reason why the photon's propagator cannot be relaxed lies in the fact that the photon's propagator is closed related to the photon-photon-$G$ vertex, so that for consistency one cannot modify the propagator solely without changing the vertex accordingly. As usual, with Feynman parametrization, we can complete the square in the denominator of the loop integral as follows
\begin{eqnarray}\label{squared}
	&&\frac{1}{[l^2-m_\ell^2] (l-k_2)^2 [(l-k_1)^2 -m_G^2]^n} \nonumber\\
	&=& \frac{\Gamma(n+2)}{\Gamma(n)} \int dx dy \frac{1}{[(l-xk_1-yk_2)^2 -\Delta_{b_1}]^{n+2}}
\end{eqnarray}
with 
\begin{eqnarray}
	\Delta_{b_1} \equiv (1-x-y)^2 m_\ell^2 -xy q^2 + x m_G^2\,.
\end{eqnarray}
As a result, we can shift the loop momentum as $l\to l+xk_1+yk_2$ in the loop integral. Then we can expand it and obtain the following relevant terms of interest
\begin{eqnarray}\label{NumB1}
	&&-\frac{16l^2}{3m_G^4 } \left\{ 2(2y-1)\slashed{l}l^\mu (k_1\cdot l) (k_2\cdot l) +4x\slashed{l} l^\mu (k_1\cdot l)^2 +\slashed{l}l^2 (k_1\cdot l) [(x-2)k_1^\mu + (1+y)k_2^\mu] \right. \nonumber\\
	&& + l^2 \big[ -l^\mu (k_1 \cdot l)\slashed{k}_2 + 2 l^\mu \slashed{k}_1 (k_2 \cdot l) -l^2 k_2^\mu \slashed{k}_1 -y l^\mu \slashed{l}\slashed{k}_2 \slashed{k}_1 + y l^\mu (k_1\cdot l) \slashed{k}_2 -4 y l^\mu \slashed{k}_1 (k_2\cdot l) \nonumber\\
	&& - y l^2 k_2^\mu \slashed{k}_1 + l^\mu \slashed{k}_1 (k_1\cdot l) + 2 l^2 k_1^\mu \slashed{k}_1 - 2 l^2 m_\ell k_1^\mu - m_\ell l^\mu (k_1 \cdot l) + m_\ell x l^\mu \slashed{k}_1 \slashed{l} + x m_\ell l^2 k_1^\mu \nonumber\\
	&& + 4xm_\ell l^\mu (k_1 \cdot l) + \xi (l\cdot q) \gamma^\mu \slashed{l}\slashed{k}_1 -  l^\mu \slashed{q} \slashed{l} \slashed{k}_1 +  l^\mu (k_1 \cdot l) \slashed{q} -5xl^\mu \slashed{k}_1 (k_1 \cdot l) -x l^2 k_1^\mu \slashed{k}_1  + m_\ell l^2 k_2^\mu \nonumber\\
	&& \left.  - 2 m_\ell l^\mu (k_2 \cdot l) +y m_\ell l^\mu \slashed{l} \slashed{k}_2 + y m_\ell l^2 k_2^\mu + 4ym_\ell l^\mu (k_2 \cdot l) - m_\ell (l\cdot q) \gamma^\mu \slashed{l} +  m_\ell l^\mu \slashed{q} \slashed{l}\big] \right\}\,.
\end{eqnarray}
According to the rules of loop regularization, we can transform the loop integral of $(b_1)$ into the sum of irreducible loop integrals which are define as follows
\begin{eqnarray}\label{IrreI}
	I_{-2m} &\equiv& \int_l \frac{1}{(l^2-M^2)^{m+2}}\,,\nonumber\\
	I_{-2m}^{\mu\nu} &\equiv& \int_l \frac{l^\mu l^\nu}{(l^2-M^2)^{m+3}}\,,\nonumber\\
	I_{-2m}^{\mu\nu\rho\sigma} &\equiv & \int_l \frac{l^\mu l^\nu l^\rho l^\sigma}{(l^2 -M^2)^{m+4}}\,,
\end{eqnarray}
where the subscripts on the left-hand side refer to the mass dimensions of loop integrals with $m \in \mathbb{Z}$. In particular, when these integrals are divergent, then the subscripts stand for the degrees of UV divergences. 
With the prescription of loop regularization,
it is easy to prove the following consistency conditions for the regularized integrals:
\begin{eqnarray}\label{IdentityLR}
	I^{\mu\nu}_{-2m} &=& \frac{g^{\mu\nu}}{2(2+m)} I_{-2m}\,,\nonumber\\
	I^{\mu\nu\rho\sigma}_{-2m} &=& \frac{1}{4(2+m)(3+m)} (g^{\mu\nu}g^{\rho\sigma}+g^{\mu\rho}g^{\nu\sigma}+g^{\nu\rho}g^{\mu\sigma}) I_{-2m}\,.
\end{eqnarray}
For the quadratically and logarithmically divergent integrals with $m=-1$ and $m=0$ respectively, the above relations are reduced as follows
\begin{eqnarray}\label{LR02}
	I^{\mu\nu}_2 &=& \frac{1}{2} g^{\mu\nu} I_2\,, \quad   I^{\mu\nu\rho\sigma}_{2} = \frac{1}{8} (g^{\mu\nu}g^{\rho\sigma}+g^{\mu\rho}g^{\nu\sigma}+g^{\nu\rho}g^{\mu\sigma}) I_{2}\,,\nonumber\\
	I^{\mu\nu}_0 &=& \frac{1}{4} g^{\mu\nu} I_0\,, \quad   I^{\mu\nu\rho\sigma}_{0} = \frac{1}{24} (g^{\mu\nu}g^{\rho\sigma}+g^{\mu\rho}g^{\nu\sigma}+g^{\nu\rho}g^{\mu\sigma}) I_{0}\,.
\end{eqnarray}
Note that it was proven that the conditions given in Eq.~(\ref{LR02}) are enough to guarantee the gauge invariance in the renormalizable gauge theories. However, now we need to extend these relations to the more general case in Eq.~(\ref{IdentityLR}) for non-renormalizable interactions.
After tedious calculations with repeated usages of the consistency conditions, the terms relevant to the lepton $g-2$ can be reduced to  
\begin{eqnarray}
	i{\cal M}_{(b_1)} &\sim & -\frac{c_\ell c_\gamma e Q_\ell}{8\Lambda^2} \left(-\frac{16}{3m_G^4}\right) m_\ell [\bar{u}(k_2) (-i\sigma^{\mu\nu}q_\nu) u(k_1)] \frac{\Gamma(n+2)}{\Gamma(n)} \nonumber\\
	&&  \int dx dy (1-x-y)^{n-1} \left\{ \frac{4(x+y-1/2)}{4(n-1)n} + \frac{x+y-1}{2(n-1)} \right\}  I_{2(3-n)} \,,
\end{eqnarray}
where the symbol of ``$\sim$'' refers to the equality up to the leading-order divergence. In general, one cannot simply integrate over the Feynman parameters, $x$ and $y$, since the loop integral $I_{2(3-n)}$ depends on them. However, here we only focus on the leading-order divergent term which does not have any reliance on Feynman parameters. Therefore, we can put the divergent part ${\cal D}[I_{2(3-n)}]$ out of the integration over $x$ with the notation ${\cal D}[..]$ denoting the leading-order divergence of the loop integral. As a result, the amplitude contributing to the lepton $g-2$ is given by
\begin{eqnarray}\label{LRB1x}
 i{\cal M}_{(b_1)}	&\sim & A_{(b_1)}  {\cal D}[I_{2(3-n)}] \frac{\Gamma(n+2)}{\Gamma(n)} \int_0^1 dt t(1-t)^{n-1} \left\{\frac{2t-1}{2n(n-1)}+\frac{t-1}{2(n-1)}\right\} \nonumber\\
	&=& A_{(b_1)} {\cal D}[I_{2(3-n)}] \frac{1}{2n(n-1)} (1-n) = A_{(b_1)} {\cal D}[I_{2(3-n)}]\left(-\frac{1}{2n}\right) \nonumber\\
	&=& -\frac{1}{2} A_{(b_1)} {\cal D}[I_{4}]\,,
\end{eqnarray} 
where we have defined 
\begin{eqnarray}
	A_{(b_1)} \equiv -\frac{c_\ell c_\gamma e Q_\ell}{8\Lambda^2} \left(-\frac{16}{3m_G^4 \xi}\right) m_\ell [\bar{u}(k_2)\left( -i \sigma^{\mu\nu} q_\nu \right) u(k_1)]\,.
\end{eqnarray}
Here we have transformed the integration over $x$ and $y$ into that over $t = x+y$  in the first relation. Further, we have taken the limit $n\to 1$ in the last equality  as demanded by the true photon propagator. Finally, the quartically divergent integral $I_4$ can be regularized by the loop regularization with the leading-order divergent part given by
\begin{eqnarray}\label{ReQD}
	{\cal D}[I_4] = \frac{i}{16\pi^2}\frac{\Lambda^4}{2}\,.
\end{eqnarray}
By comparing the definition of $\Delta a_\ell$ in Eq.~(\ref{g2Def}),  we can extract the contribution to the muon anomalous magnetic moment as follows
\begin{eqnarray}\label{ResultB1}
	\Delta a^{(b_1)}_\ell \approx -\frac{c_\ell c_\gamma}{48\pi^2} \left(\frac{m_\ell}{\Lambda}\right)^2 \left(\frac{\Lambda}{m_G}\right)^4\,.
\end{eqnarray}

\subsection{Diagram $(b_1)$ in the General Gauge}
As a proof of gauge invariance of quantum electrodynamics, we perform the calculation in the general gauge with $\xi$ in Eq.~(\ref{ExpB1}) taken as arbitrary values. 
Following exactly the same procedure, we can firstly complete the square in the denominator as follows
\begin{eqnarray}
	&& \frac{1}{[l^2-m_\ell^2] (l-k_2)^4 [(l-k_1)^2 -m_G^2]^n} \nonumber\\
	&=& \frac{\Gamma(n+3)}{\Gamma(2)\Gamma(n)} \int dx dy y \frac{1}{[(l-xk_1-yk_2)^2-\Delta_{b_1}]^{(n+3)}}\,.
\end{eqnarray}
After the shift of loop momentum $l$, the terms in the numerator relevant to the lepton $g-2$ are given by
\begin{eqnarray}\label{NumB1p}
	&&-\frac{16l^4}{3m_G^4} \left\{ 2(3y-2)\slashed{l}l^\mu (k_1\cdot l) (k_2\cdot l) +6x\slashed{l} l^\mu (k_1\cdot l)^2 +\slashed{l}l^2 (k_1\cdot l) [(x-2)k_1^\mu + (1+y)k_2^\mu] \right. \nonumber\\
	&& + l^2 \big[ -l^\mu (k_1 \cdot l)\slashed{k}_2 + 4 l^\mu \slashed{k}_1 (k_2 \cdot l) -l^2 k_2^\mu \slashed{k}_1 -y l^\mu \slashed{l}\slashed{k}_2 \slashed{k}_1 + y l^\mu (k_1\cdot l) \slashed{k}_2 -6 y l^\mu \slashed{k}_1 (k_2\cdot l) \nonumber\\
	&& - y l^2 k_2^\mu \slashed{k}_1 + l^\mu \slashed{k}_1 (k_1\cdot l) + 2 l^2 k_1^\mu \slashed{k}_1 - 2 l^2 m_\mu k_1^\mu - m_\mu l^\mu (k_1 \cdot l) + m_\mu x l^\mu \slashed{k}_1 \slashed{l} + x m_\mu l^2 k_1^\mu \nonumber\\
	&& + 6xm_\mu l^\mu (k_1 \cdot l) + (l\cdot q) \gamma^\mu \slashed{l}\slashed{k}_1 - l^\mu \slashed{q} \slashed{l} \slashed{k}_1 + l^\mu (k_1 \cdot l) \slashed{q} -7xl^\mu \slashed{k}_1 (k_1 \cdot l) -x l^2 k_1^\mu \slashed{k}_1  + m_\mu l^2 k_2^\mu \nonumber\\
	&& \left.  - 4m_\mu l^\mu (k_2 \cdot l) +y m_\mu l^\mu \slashed{l} \slashed{k}_2 + y m_\mu l^2 k_2^\mu + 6ym_\mu l^\mu (k_2 \cdot l) - m_\mu (l\cdot q) \gamma^\mu \slashed{l} + m_\mu l^\mu \slashed{q} \slashed{l}\big] \right\}\,,
\end{eqnarray}
where we have also suppress the prefactors involving the coupling constants for simplicity.
It is obvious that the gauge parameter $\xi$ has been canceled out completely, which means that our final result is gauge-independent. With the general consistency relations in Eqs.~(\ref{IdentityLR}), the corresponding amplitude is given by
\begin{eqnarray}\label{LRB1}
	i{\cal M}_{(b_1)} &\sim & A_{(b_1)} \frac{\Gamma(n+3)}{\Gamma(n)\Gamma(2)} \int dx dy y(1-x-y)^{n-1} \left\{ \frac{2(3x+3y-2)}{4(n-1)n} + \frac{x+y-1}{2(n-1)} \right\}  I_{2(3-n)}\nonumber\\
	&=& A_{(b_1)} {\cal D}[I_{2(3-n)}]  \frac{1}{4n(n-1)}  \left[6- 2(n+2)\right] \nonumber\\ 
	&\sim& -\frac{1}{2} A_{(b_1)} {\cal D}[I_{4}] 
	\sim \frac{c_\ell c_\gamma e Q_\ell}{96\pi^2\Lambda^2} m_\ell (i\sigma^{\mu\nu}q_\nu) \left(\frac{\Lambda}{m_G}\right)^4 \,,
\end{eqnarray}   
where, in the last line, we have taken the limit $n\to 1$ to recover the original one-loop integral of $(b_1)$. The contribution to the lepton $g-2$ is given by 
\begin{eqnarray}
	\Delta a^{(b_1)}_\ell \approx -\frac{c_\ell c_\gamma}{48\pi^2} \left(\frac{m_\ell}{\Lambda}\right)^2 \left(\frac{\Lambda}{m_G}\right)^4\,,
\end{eqnarray}  
which agrees with Eq.~(\ref{ResultB1}) for the fixed Feynman-'t Hooft gauge. Therefore, we have explicitly shown that our result respects the gauge invariance,
which guarantees the consistency of our calculation. 


\subsection{Diagram $(b_2)$}
Similar to the diagram $(b_1)$, the amplitude of the Feynman diagram $(b_2)$ is given by
\begin{eqnarray}
	i{\cal M}_{(b_2)} &=& -\frac{c_\ell c_\gamma e Q_\ell}{8\Lambda^2} \frac{\bar{u}(k_2)\left\{ \gamma^\rho (l+k_2)^\sigma + \gamma^\sigma (l+k_2)^\rho -2 \eta^{\rho\sigma} (\slashed{k}_2 + \slashed{l}-2m_\ell) \right\} (\slashed{l}+m_\ell)\gamma^\nu u(k_1)}{(l^2-m_\ell^2)[(l-k_2)^2-m_G^2](l-k_1)^2} \nonumber\\
	&& B_{\alpha\beta, \rho\sigma}(l-k_2) [q\cdot (l-k_1) C_{\alpha\beta, \mu\kappa}+D_{\alpha\beta,\mu\kappa}(q,l-k_1)+\xi^{-1} E_{\alpha\beta,\mu\kappa}(q,l-k_1)]A_{\nu\kappa}(l-k_1)\,.\nonumber\\
\end{eqnarray}
Following the same treatment as for the diagram $(b_1)$,  the contribution from $(b_2)$ to the lepton $g-2$ is given by
\begin{eqnarray}
	\Delta a_\ell^{(b_2)} \approx -\frac{c_\ell c_\gamma}{48\pi^2} \left(\frac{m_\ell}{\Lambda}\right)^2 \left(\frac{\Lambda}{m_G}\right)^4\,.
\end{eqnarray}

\subsection{Diagram $(c)$}
According to the Feynman rules, the amplitude of the diagram $(c)$ is given by
\begin{eqnarray}\label{ExpC}
	&& i{\cal M}_{(c)} = \frac{ic_\ell e Q_\ell}{2\Lambda} \int\frac{d^4l}{(2\pi)^4} \bar{u}(k_2) (C^{\rho \sigma, \nu\tau}-\eta^{\rho\sigma} \eta^{\nu\tau})\gamma_\tau u(k_1) \frac{(-i)A_{\kappa\nu}(l+q)}{(l+q)^2} \left(-\frac{ic_\gamma}{\Lambda} \right) \nonumber\\
	&&\left[ -q\cdot (l+q) C^{\alpha \beta, \mu\kappa} + D^{\alpha\beta, \mu\kappa}\big(q,-(l+q)\big) +\xi^{-1} E^{\alpha \beta, \mu\kappa} \big( q, -(l+q) \big) \right] \frac{iB_{\alpha\beta, \rho\sigma}(l)/2}{l^2-m_G^2} \,, 
\end{eqnarray}
where the tonsorial factor  $A_{\mu\nu}(p)$ and $B_{\mu\nu\,\rho\sigma} (k)$ are defined in Eqs.~(\ref{PropGamma}) and (\ref{PropG}). 
Explicit calculations show that, by contracting all Lorentz indices and using the on-shell condition for the external photon in Eq.~(\ref{ExpC}), the remaining terms are all proportional to $\gamma^\mu$ so that this diagram cannot contribute to the magnetic dipole of leptons.

\subsection{Diagrams $(d_1)$ and $(d_2)$}
In this subsection, we present the computational details of Feynman diagrams of $(d_1)$ and $(d_2)$. Since they are symmetric, it is expected that both of them give the same result. Straightforward computations confirm this conclusion. Therefore, here we only show the necessary steps for the calculation of diagram $(d_1)$ in the following.

The amplitude of $(d_1)$ is given by
\begin{eqnarray}
	i{\cal M}_{(d_1)} &=&  \int\frac{d^4 l}{(2\pi)^4} \bar{u}(k_2) \left(-i\frac{c_\ell}{4\Lambda} \right) \left[ \gamma^\rho (l + 2k_2)^\sigma + \gamma^\sigma (l+2k_2)^\rho -2\eta^{\rho\sigma} (\slashed{l}+2\slashed{k}_2 - 2m_\ell) \right]  \nonumber\\
	&& \frac{i}{\slashed{l}+\slashed{k}_2 - m_\ell} \left[\frac{ieQ_\ell c_\ell}{2\Lambda} \big( C^{\alpha \beta, \mu\nu} -\eta^{\alpha\beta} \eta^{\mu\nu} \big) \gamma_\nu \right] u(k_1) \left( \frac{iB_{\alpha \beta, \rho\sigma} (l)/2}{ l^2-m_G^2} \right)\nonumber\\
	&=& -\frac{eQ_\ell c_\ell^2}{16\Lambda^2} \int \frac{d^4 l}{(2\pi)^4} \big(C^{\alpha\beta, \mu\nu} - \eta^{\alpha\beta} \eta^{\mu\nu} \big) B_{\alpha\beta,\rho\sigma} (l) \nonumber\\
	&& \frac{\bar{u}(k_2) \left[ \gamma^\rho (l+2k_2)^\sigma + \gamma^\sigma (l+2k_2)^\rho -\eta^{\rho\sigma} (\slashed{l}+2\slashed{k}_2 -2 m_\ell) \right]  \left[ \slashed{l} + \slashed{k}_2 + m_\ell \right] \gamma_\nu u(k_1)}{[(l+k_2)^2 - m_\ell^2] (l^2-m_G^2)}\,.
\end{eqnarray} 
Like the situation when computing the diagram $(b_1)$, the application of the consistency conditions in Eq.~(\ref{IdentityLR}) would cause the denominator to become 0. In order to cure this unphysical pathology, we shall follow Eq.~(\ref{PropGN}) to relax the power index of the spin-2 particle's propagator from 1 to $n$, and take the limit $n\to 1$ in the final expression. 
With this prescription, we complete the square in the denominator as follows
\begin{eqnarray}
	\frac{1}{[(l+k_2)^2-m_\ell^2](l^2-m_G^2)^n} = \frac{\Gamma(n+1)}{\Gamma(n)} \int^1_0 dx \frac{1}{[(l+xk_2^2)-\Delta_d]^{n+1}}\,, 
\end{eqnarray}
where 
\begin{eqnarray}
	\Delta_d = x m_\ell^2 +(1-x)m_G^2 - x(1-x)k_2^2\,.
\end{eqnarray}
Now we shift the loop momentum $l \to l-x k_2$ so that the numerator in the loop integral can be transformed into
\begin{eqnarray}
	N &=& \bar{u}(k_2) \left\{ \gamma^\rho [l^\sigma + (2-x)k_2^\sigma] + \gamma^\sigma [l^\rho + (2-x)k_2^\rho] - 2\eta^{\rho\sigma} [\slashed{l}+(2-x)\slashed{k}_2 -2m_\ell] \right\} \nonumber\\
	&& [\slashed{l}+(1-x)\slashed{k}_2 + m_\ell] \gamma_\nu u(k_1) (C^{\alpha\beta, \mu\nu}-\eta^{\alpha\beta}\eta^{\mu\nu}) B_{\alpha\beta,\rho\sigma} (l-xk_2)\,,
\end{eqnarray}
where we have omitted the prefactors for simplicity. 
By contracting the Lorentz indices and simplifying the expression with the on-shell conditions of external particles, we find that the terms which potentially contribute at leading order to the lepton magnetic dipole operator are given by
\begin{eqnarray}
	N^\prime = -\frac{32}{3 m_G^4} \left\{ l^2[xm_\ell l^\mu \slashed{l} \slashed{k}_2 - x m_\ell l^\mu (k_2\cdot l) + x k^\mu_2 \slashed{l}(k_2\cdot l) ]  + 2(x-1) l^{\mu} (k_2\cdot l)^2 \slashed{l}\right\}\,.
\end{eqnarray} 

Hence, when written in terms of the irreducible loop integrals, the above amplitude can be expressed as follows
\begin{eqnarray}
	i{\cal M}_{(d_1)} &\sim& -\frac{eQ_\ell c_\ell^2}{16\Lambda^2} \left(-\frac{32}{3m_G^4}\right) \frac{\Gamma(n+1)}{\Gamma(n)} \int^1_0 dx \left[ x m_\ell I^{\mu\nu}_{2(3-n)} \gamma_\nu \slashed{k}_2 + 2(x-1) I^{\mu\nu\rho\sigma}_{2(3-n)} \gamma_\nu k_{2\rho} k_{2\sigma} 
	\right] \nonumber\\
	&\sim& \frac{eQ_\ell c_\ell^2}{16\Lambda^2} \left(\frac{32 m_\ell}{3m_G^4}\right) \frac{\Gamma(n+1)}{\Gamma(n)} \int^1_0 dx \left[ \frac{x}{2(n-1)} \gamma^\mu \slashed{k}_2 - \frac{(1-x)}{n(n-1)} k^\mu \right] I_{2(3-n)} \nonumber\\
	&\sim &  \frac{eQ_\ell c_\ell^2}{16\Lambda^2} \left(\frac{32 m_\ell}{3m_G^4}\right) \left(-i\sigma^{\mu\nu}q_\nu \right) \frac{\Gamma(n+1)}{\Gamma(n)} \int^1_0 dx \left[ \frac{x}{2(n-1)} - \frac{(1-x)}{2n(n-1)}\right] I_{2(3-n)} \,,
\end{eqnarray}
where we have applied the consistency conditions in Eq.~(\ref{IdentityLR}) and have only kept terms that give rise to the lepton $g-2$. Note that the leading-order divergence denoted as ${\cal D}[I_{2(3-n)}]$ is independent of the Feynman parameter $x$, so that we can take it our of the integration over $x$. 
As a result, the dominant amplitude proportional to $(i\sigma^{\mu\nu}q_\nu)$ is given by
\begin{eqnarray}
	i{\cal M}_{(d_1)} &\sim& \frac{eQ_\ell c_\ell^2}{16\Lambda^2} \left(\frac{32 m_\ell}{3m_G^4}\right) \left(-i\sigma^{\mu\nu}q_\nu \right) \frac{\Gamma(n+1)}{\Gamma(n)} {\cal D}[I_{2(3-n)}] \left[ \frac{1}{4(n-1)} - \frac{1}{4n(n-1)} \right] \nonumber\\
	&=& \frac{eQ_\ell c_\ell^2}{16\Lambda^2} \left(\frac{32 m_\ell}{3m_G^4}\right) \left(-i\sigma^{\mu\nu}q_\nu \right) \frac{\Gamma(n+1)}{\Gamma(n)} {\cal D}[I_{2(3-n)}] \frac{1}{4n} \nonumber\\
	&\sim &  \frac{eQ_\ell c_\ell^2}{16\Lambda^2} \left(\frac{8 m_\ell}{3m_G^4}\right) \left(-i\sigma^{\mu\nu}q_\nu \right)  {\cal D}[I_{4}]\,,
\end{eqnarray}
where in the last line we have taken the limit $n\to 1$. 
By comparing with the definition of $\Delta a_\ell$ in Eq.~(\ref{g2Def}), the contribution from diagram $(d_1)$ to the lepton $g-2$ is given by
\begin{eqnarray}
	\Delta a_\ell^{(d_1)} = \frac{c_\ell^2}{96\pi^2} \left(\frac{m_\ell}{\Lambda}\right)^2 \left(\frac{\Lambda}{m_G}\right)^4\,,
\end{eqnarray}
where we have used the expression of the top divergence in Eq.~(\ref{ReQD}) obtained by the loop regularization.

By explicit calculation of the diagram $(d_2)$,  we can prove that it gives exactly the same leading-order contribution to the lepton $g-2$ as $(d_1)$: $\Delta a_\ell^{(d_2)} = \Delta a_\ell^{(d_1)}$. 

\subsection{Total Contribution to $\Delta a_\mu$}
By summing up all of the contributions above to the lepton $g-2$, we can obtain the total one-loop contribution induced by the massive spin-2 field $G$ as follows
\begin{eqnarray}
	\Delta a_\ell^G = \left(\frac{m_\ell}{\Lambda}\right)^2 \left(\frac{\Lambda}{m_G^4}\right) \left( \frac{c_\ell^2}{48\pi^2}-\frac{c_\ell c_\gamma}{24\pi^2}\right)\,.
\end{eqnarray}
We shall apply this analytic expression in our numerical investigation over the parameter space in the main text.


\end{document}